\documentclass[usenatbib,twocolumn]{aastex631}

\usepackage{graphicx}	% Including figure files
\usepackage{amsmath}	% Advanced maths commands
\usepackage{xcolor}
\usepackage{booktabs}
\usepackage{threeparttable}
\usepackage{comment}
\usepackage{newtxtext,newtxmath}
\usepackage{hyperref}
\usepackage{array}

\usepackage[T1]{fontenc}

\DeclareRobustCommand{\VAN}[3]{#2}
\let\VANthebibliography\thebibliography
\def\thebibliography{\DeclareRobustCommand{\VAN}[3]{##3}\VANthebibliography}

\begin{document}

% \title{SEISTRON: Stellar Evolution and Interior STructures with
% vaRiability and Optimized Neural networks}
\title[Benchmarking Asteroseismic ML Emulators]{Benchmarking Machine Learning Emulators of Stellar Evolution for Precision Asteroseismology}

\correspondingauthor{Naomi Gluck}
\affiliation{Department of Physics, Yale University, New Haven, CT 06510, USA}
\email{naomi.gluck@yale.edu}

\author[0009-0003-4100-4710]{Naomi Gluck}
\affiliation{Department of Physics, Yale University, New Haven, CT 06510, USA}

\author[0000-0003-4456-4863]{Earl P. Bellinger}
\affiliation{Department of Astronomy, Yale University, New Haven, CT 06510, USA}
\affiliation{Institute for Foundations of Data Science, Yale University, New Haven, CT 06510, USA}

\author[0000-0002-1001-1235]{Yan Liang}
\affiliation{Department of Astronomy, Yale University, New Haven, CT 06510, USA}

\author[0000-0002-5794-4286]{Ebraheem Farag}
\affiliation{Department of Astronomy, Yale University, New Haven, CT 06510, USA}

\author[0000-0003-2657-3889]{Nicholas Saunders}
\affiliation{Department of Astronomy, Yale University, New Haven, CT 06510, USA}

\author[0009-0001-6503-9841]{Selim Kalici}
\affiliation{Department of Astronomy, Yale University, New Haven, CT 06510, USA}

\author[0000-0001-8722-1436]{Christopher J. Lindsay}
\affiliation{Department of Astronomy, Yale University, New Haven, CT 06510, USA}

\author[0000-0002-6163-3472]{Sarbani Basu}
\affiliation{Department of Astronomy, Yale University, New Haven, CT 06510, USA}

%\email{naomi.gluck@yale.edu}



\begin{abstract}
Fast and accurate stellar evolution emulators---surrogate models that approximate expensive simulation outputs with machine learning (ML)---are powerful tools for modern stellar characterization, hierarchical inference, and population synthesis. We analyze the grid density required for reliable emulation by training ML algorithms on main-sequence models with masses $M\in[0.7,1.2]\,M_{\odot}$. This range is challenging to emulate due to rapidly varying evolutionary behavior caused by the radiative-to-convective core transition, as well as the requirement to match the part-per-thousand seismic precision that has been delivered for such stars from the NASA \textit{Kepler} mission. Generating grids from analytical models, as well as \texttt{MESA}, \texttt{YREC}, \texttt{MIST}, and \texttt{ASTEC}, we compare linear interpolation, k-nearest neighbors, random forests, and neural networks (NNs) in interpolating the stellar observables $T_{\rm eff}$, $L$, $\Delta\nu$, and $\nu_{\rm max}$. 
While NNs outperform other methods, sparse grids induce localized failures in the core-transition region, resulting in unstable derivatives, ensemble disagreement, and fragmented posterior distributions during inference. Performance gains from denser grids are non-uniform, suggesting that adaptive grid generation should be favored over uniform refinement. Finally, we show that NN ensembles allow for localized uncertainty propagation, more accurately reflecting emulator reliability across parameter space than global uncertainty estimates.
As we consider only the two-dimensional case of varying only stellar mass and age along the main sequence, these results represent a lower bound on the challenge in emulating stellar evolution simulations for precision asteroseismology.
\end{abstract}

% \begin{keywords}
% accretion, accretion disks; Galaxy: evolution, formation, halo; methods: analytical
% \end{keywords}

\section{Introduction}
\label{sec:intro}

Asteroseismology---the study of stellar oscillations---has transformed our ability to infer precise stellar properties, including mass, radius, and age across large stellar populations \citep{aerts_2010}. Extracting these properties to characterize stars from observations requires stellar evolution models that reproduce the internal structure and hence oscillation frequencies of stars spanning a broad range of masses, metallicities, and evolutionary stages, and span the theoretical uncertainties underpinning the simulations. 

Large-scale surveys, and particularly the NASA \textit{Kepler} \citep{Borucki_kepler_2016} and \textit{TESS} \citep{Ricker_tess_2022} missions, have delivered high-quality asteroseismic measurements that enable detailed study of the internal stellar structure \citep[e.g.,][]{2017ApJ...851...80B, 2019ApJ...885..143B, 2024ApJ...961..198B, 2025ApJ...987...97B} and greatly improve inferences of fundamental stellar properties \citep[e.g.,][]{Quirion_2010,bellinger_2016,Nielsen_2020, Li_2023, theodoridis_2025}. Such inferences support a wide variety of endeavors, such as characterizing exoplanetary systems \citep[e.g.,][]{2014ApJ...782...14V, 2015ApJ...799..170C, 2015MNRAS.452.2127S, 2019A&A...622A.130B} and inferring detailed information about the formation and evolution of the Milky Way \citep[e.g.,][]{miglio_2013,Pinsonneault_2014, sharma_2016, Stokholm_2023,Lindsay_2025}. The number of observed asteroseismic targets is expected to increase more than tenfold in the coming years \citep{huber_2020} with the coming launch of the ESA PLATO \citep{rauer2024platomission} and NASA \textit{Roman} \citep{Weiss_2025} spacecraft.

State-of-the-art stellar evolution codes like \texttt{MESA} \citep{paxton_2011,Paxton2013,Paxton2015,Paxton2018, Paxton2019,Jermyn2023}, \texttt{ASTEC} \citep{Christensen_Dalsgaard_2007}, \texttt{PARSEC} \citep{Bressan_2012, Nguyen_2025}, \texttt{YREC} \citep{Demarque_2008, Pinsonneault_2025, Pinsonneault_2026}, along with many others, have become essential tools for stellar astrophysics and galactic archaeology \citep[see][for review]{basu_hekker_2020}. Stellar evolution simulations output one evolutionary \textit{track} per star, consisting of many \textit{models}, or snapshots, used to record changes in fundamental parameters throughout a star's evolution. In an ideal case with infinite computing resources and simulations that run instantaneously, a simulated version of every observed star could be generated for analysis. However, running such simulations with high precision is impossible due to the high computational cost. To mitigate this, tracks are stacked to create a \textit{grid}, which is then interpolated using numerical or machine-learning methods as efficient emulators to provide continuous coverage of a discrete parameter space. 

Several widely-used pre-computed grids are publicly available, including \texttt{BaSTI} \citep{Pietrinferni_2004, Hidalgo_2018}, Geneva Stellar Evolution Tracks \citep{ekstrom_2012}, \texttt{YaPSI} \citep{spada_2017_YaPSI}, and \texttt{MIST} \citep{choi_2016, dotter_2016}, among others. These grids and similar ones have previously been interpolated with various methods: linear and cubic spline interpolation \citep{clara_2025}, random forest regression \citep{bellinger_2016}, Gaussian process regression \citep{Li_2022}, artificial neural networks \citep[ANNs,][]{lyttle_2021, Maltsev_2024,saunders_2024, scutt_2026}, and normalizing flows \citep{hon_2024, stone-martinez_2025}, among others.

More specifically, emulators, or interpolators, of these stellar grids facilitate the inference of stellar masses, radii, and ages from asteroseismic measurements by providing a differentiable, probabilistic mapping that accounts for model uncertainty. For solar-like oscillators (stars with an outer convective envelope), the mapping between observed quantities and fundamental stellar properties includes two global asteroseismic quantities derived from stochastically excited acoustic pressure modes: $\Delta\nu$ (the average spacing between consecutive overtones of the same spherical degree) and $\nu_{\rm max}$ (the frequency at maximum oscillation power), which scale with the mass and radius (Equations \ref{eq:Delta_nu} and \ref{eq:numax}, respectively). Based on these relations, the uncertainties of both $\Delta\nu$ and $\nu_{\rm max}$ propagate directly into the characterization of the stellar mass and radius, and by extension, the age \citep[e.g.,][]{ulrich_1986, gough_1987, cd_1988, Kjeldsen_1995, stello_2009, basu_2010, Creevey_2012, stancliffe_2016, bellinger_2019, Bellinger_age_mass_radii_2019}.

Beyond refinement of single-star characterization, the ability to rapidly propagate these probabilistic mappings facilitates hierarchical modeling \citep[e.g.,][]{lyttle_2021}. In this statistical framework, individual stellar parameters are treated as draws from a shared population distribution, allowing for the simultaneous inference of both individual properties and population-level parameters that govern the ensemble. Furthermore, these efficient surrogates would enable constraints on additional physical processes \citep[such as reaction rates, e.g.,][]{Bellinger_2022}, and serve as the backbone for both single and binary star evolution \citep{Srivastava_2024, Teng_2025} within population synthesis codes like \texttt{POSYDON} \citep{Fragos_2023_Posydon_I, andrews2025posydonversion2population}.

The ability of the aforementioned advancements in stellar characterization is highly dependent on the underlying grid. If the grid of tracks is not sufficiently dense, accurate and precise interpolation becomes challenging. This is especially true in regions of rapidly changing physics, where interpolation errors can exceed observational uncertainties and bias inferred stellar parameters. However, this is rarely thoroughly checked. To mitigate this, it is essential that the uncertainty of the interpolation is rigorously quantified and propagated when using an interpolated grid to characterize an observed system.

\begin{figure*}
    \centering
    \includegraphics[width=\linewidth]{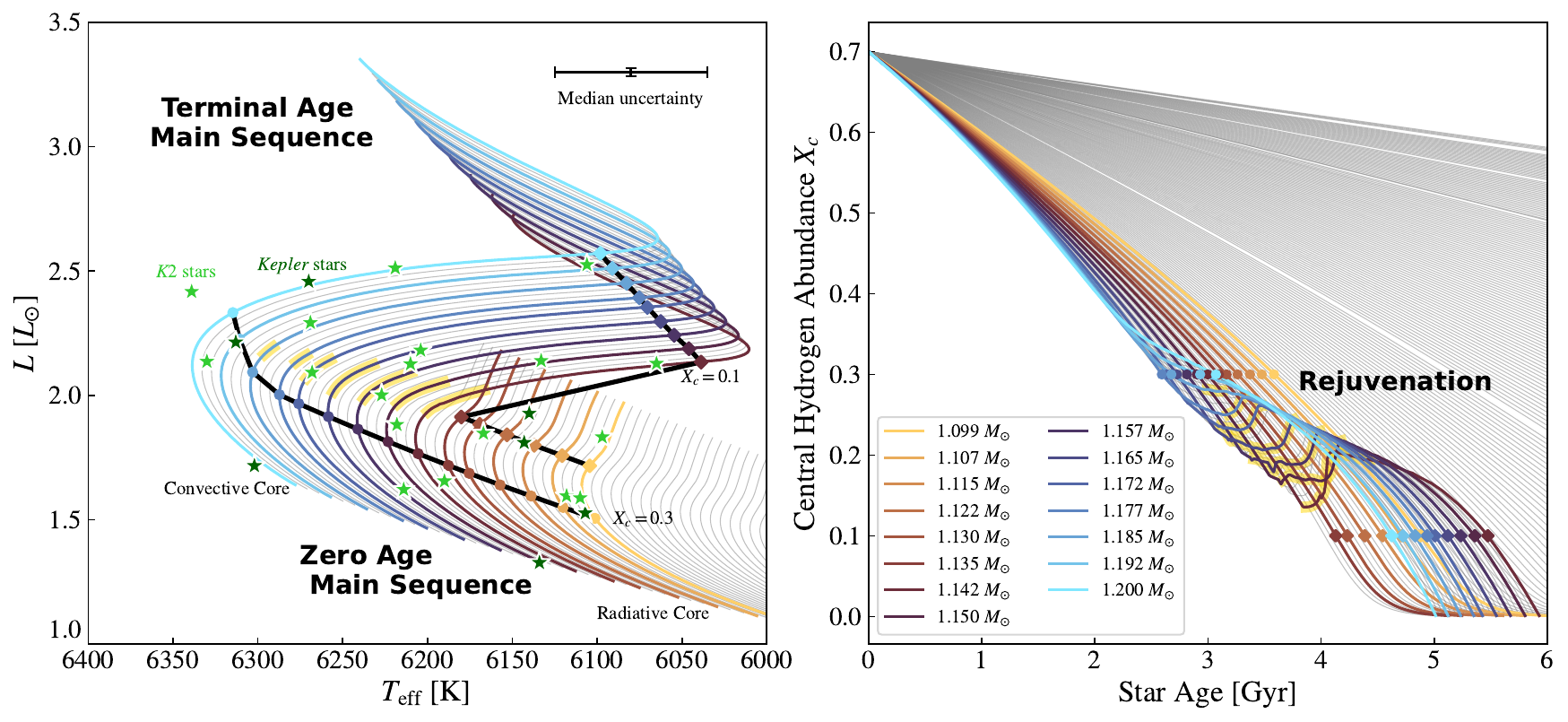}
    \caption{\textit{Left}: HR diagram showing 15 example \texttt{MESA} stellar evolution tracks for masses $M \in [1.099,1.2]\ M_{\odot}$ with initial metallicity $Z_0=0.02$ that evolve to the end of the MS (full grid in the background). These models illustrate the transition region between stars that are born with radiative cores and those born with convective cores. We overplot the \textit{Kepler}-LEGACY and \textit{K2} stars with their uncertainties (luminosity uncertainties from \textit{Gaia}) that fit within this narrow region. We show the combined median uncertainty for both surveys. \textit{Right}: The evolution of the central hydrogen fractional abundance $X_c$ as a function of stellar age for the same set of stars is shown in the right panel. Stars with $M \in [1.145,1.17]\ M_{\odot}$ are rejuvenated due to the onset of convection partway through the main sequence. The non-monotonicity is a justification for our focus on this mass range and evolutionary phase.}
    \label{fig:hr_tracks_Xc_abundance}
\end{figure*}

Figure~\ref{fig:hr_tracks_Xc_abundance} highlights the ``difficult" transition region for 15 representative tracks from a \texttt{MESA}-generated grid in which only the initial mass and age are varied (see Sec.~\ref{sec:mesa_grid} for additional details). On the main sequence (MS), stars below \(\sim 1.145\,M_{\odot}\) retain radiative cores, whereas stars above \(\sim 1.185\,M_{\odot}\) have convective cores. By contrast, stars in the range \(1.145 \le M/M_{\odot} \le 1.185\) begin with radiative cores but later develop convective cores. The mass values that define this transition region depend on the physics implementations within the simulation and the initial conditions used to generate the tracks. 

As the core temperature rises, CNO burning becomes increasingly important because its energy generation rate is much more temperature-sensitive than that of the proton-proton (pp) chain; it steepens the central radiative temperature gradient and eventually drives core convection. The resulting convective mixing causes \textit{rejuvenation}, as this process brings fresh hydrogen into the core, and hence, the central hydrogen abundance, \(X_c\), is no longer strictly monotonic. This causes the standard ``equivalent evolutionary phase'' \citep[EEP,][]{prather_1976} construction based on monotonic central-hydrogen depletion to fail in this mass range. Stars that undergo this rejuvenation then go on to have the characteristic hook feature in the HR diagram at the end of the main sequence. 

Rejuvenation also produces a local non-monotonicity in the mass--lifetime relation: normally, higher mass stars evolve faster, but the higher-mass \(1.145\,M_{\odot}\) model remains on the main sequence longer than the lower-mass \(1.13\,M_{\odot}\) model. This transition underlies a persistent debate over the internal structure of $\alpha$~Centauri~A and whether it has a convective core \citep{brandao_2014, bazot_2016, nsamba_2018}. The nature of the core in this mass range is made even more challenging by uncertainties in convective boundary mixing (which extends the size of the core) and in notable nuclear reaction rates such as ${}^{14}\mathrm{N}(p,\gamma){}^{15}\mathrm{O}$, the slowest CNO cycle rate, which can heavily influence the size of main-sequence convective cores. Such uncertainties in principle should be propagated when characterizing stars in this mass range, thus adding more dimensions to the simulations to be emulated. 

Before using any generated grid for interpolation, regardless of the scientific goal, one should ask if the grid is of sufficient density and smoothness for interpolation. While internal numerical convergence (e.g., \texttt{MESA} consistently simulates the same evolutionary track for the same initial conditions) is necessary, it does not guarantee global smoothness across a grid. Small discontinuities or nonlinearities between neighboring tracks can induce interpolation errors that exceed current observational uncertainties.

In this work, we showcase a pipeline for emulating grids of stellar tracks with various ML methods, quantifying success within the context of observational survey capabilities, and propagating emulator uncertainty. We focus on the mass range $M\in[0.7,1.2]\ M_{\odot}$, as it marks the transition on the MS from a radiative to a convective core. This transition is both numerically challenging to model and numerically challenging to interpolate.

We systematically show the impact of having an insufficiently dense grid, and how increasing the amount of training data allows for the random uncertainties from the interpolation to be subdominant to the observational uncertainties. We also show how the uncertainty is propagated from asteroseismic quantities to fundamental stellar parameters, allowing for an asteroseismic precision threshold determined by the science case that must be met by the interpolator. Furthermore, we present a comparison of several algorithms that are widely used for interpolating stellar evolution simulations: linear interpolation, k-nearest neighbors, random forests, and artificial neural networks. We provide a framework for determining sufficient grid density rather than a single number, as this should be individually determined for each science case.

The remainder of this paper is organized as follows. Section~\ref{sec:methods} presents the grid specifications (\S~\ref{sec:grid_generation}), the machine learning methods for interpolation (\S~\ref{sec:interpolation}), and uncertainty calculations (\S~\ref{sec:uncertainty}). We quantify the success of various ML methods on an analytical grid in Section~\ref{sec:quantify_ml_method}, explore the implications of grid density in Section~\ref{sec:grid_density}, quantify and propagate the uncertainty in Section~\ref{sec:uq}, and interpolate precomputed grids in Section~\ref{sec:other_grids}. We discuss previous interpolation efforts in Section~\ref{sec:previous_works}, and potential pipeline improvements in Section~\ref{sec:improvements}. We summarize our findings, discuss implications for higher-dimensional emulation and individual frequency interpolation, and offer broader conclusions for the field in Section~\ref{sec:conclusions}.

\section{Methods}
\label{sec:methods}

\subsection{Grid Generation and Preprocessing}
\label{sec:grid_generation}

We employ two grids in this work: a numerical grid we generate using \texttt{MESA}, and an analytical grid we generate using the \citep{Hurley_2000} single-star evolution fitting formulae. We use the analytical grid to complement our analysis because it isolates numerical artifacts that can arise in stellar evolution simulations. 

Each generated grid in this work contains three categories of variables: input interpolation parameters, output interpolation parameters, and track number. Figure~\ref{fig:hr_tracks_Xc_abundance} shows 15 representative tracks of a simple 200-track \texttt{MESA} grid varying only initial mass and age. The top left panel of Figure~\ref{fig:hr_other_grids} in Appendix~\ref{sec:model_specs} shows 50 representative tracks of a 2000-track analytical grid following \cite{Hurley_2000}.

Input interpolation parameters include varied initial mass ($M \in [0.7,1.2]\ M_{\odot}$ of various step sizes) and age. Age, as opposed to the stellar radius or the central Hydrogen abundance, is used due to its monotonicity along the track, and is thus preferred when building a well-defined, non-injective mapping. For the most straightforward interpolation case, we keep constant values of initial metallicity $Z_{0}$, helium mass fraction $Y_{0}$, mixing-length parameter $\alpha_{\rm MLT}$, core overshoot $f_{\rm ov, core}$, and shell overshoot $f_{\rm ov, shell}$. Output parameters include the effective temperature $T_{\rm eff}$, luminosity $L$, $\Delta \nu$, and $\nu_{\rm max}$, where $\Delta\nu$ and $\nu_{\rm max}$ are calculated using the following scaling relations:
\begin{equation}\label{eq:Delta_nu}
    \Delta \nu \approx \left(\frac{M}{M_{\odot}}\right)^{1/2} \left(\frac{R}{R_{\odot}}\right)^{-3/2} \Delta\nu_{\odot}
\end{equation}
\begin{equation}\label{eq:numax}
    \nu_{\rm max} \approx \frac{M}{M_{\odot}} \left( \frac{R}{R_{\odot}}\right)^{-2} \left( \frac{T_{\rm eff}}{T_{\rm eff, \odot}} \right)^{-1/2} \nu_{\rm max, \odot}
\end{equation}
where $T_{\rm eff, \odot}=5777\ {\rm K}$, $\Delta\nu_{\odot}=135.1\ \mu{\rm Hz}$, and $\nu_{\rm max, \odot}= 3090\ \mu{\rm Hz}$
\citep[see e.g.,][for review]{garcia_2019}. We note that these scaling relations are approximations, and a more accurate calculation of $\Delta \nu$ and $\nu_{\rm max}$ uses the individual mode frequencies. 

We also include a unique track identifier used to keep full tracks together when splitting the full grid into training, validation, and testing data. Prior to any interpolation, we log-scale only the output quantities ($T_{\rm eff}$, $L$, $\Delta\nu$, and $\nu_{\rm max}$), and then apply individual standard scalers (fit to only the training data) to each input and output parameter within the entire grid. After training, we obtain predictions and residuals from the held-out test set, which are computed in the described transformed space and converted back to physical units. A linear age spacing (in Gyr) is used rather than a logarithmic spacing to avoid spurious extrapolation when predicting parameters at very young ages. 

\subsubsection{Analytical Grids}

To determine how many tracks are needed to train the algorithms in an idealized case (without numerical noise from a simulation), we generate an analytical grid following \cite{Hurley_2000} using \texttt{Starpasta}\footnote{Python implementation: \href{https://github.com/hersfeldtn/Starpasta.git}{https://github.com/hersfeldtn/Starpasta.git}}. These analytical fitting formulae for stellar evolution are based on stellar models computed in \cite{pols_1998} for masses $M \in [0.5, 50]\ M_{\odot}$, metallicities $Z = [0.0001, 0.0003, 0.001, 0.004, 0.01, 0.02, 0.03]$, and $\alpha_{\rm MLT}=2.0$. For each $Z$, they compute 25 tracks varying in mass with an approximate spacing of $0.1$ in $\log(M)$, with four additional models within $M\in [0.8, 2.0]\ M_{\odot}$ due to rapidly changing MS path shapes in this range. They additionally consider overshoot with a free parameter $\delta_{\rm ov}=0.12$ (different from the commonly used definition that relates the overshooting distance to the pressure scale height) to obtain formulae that are within 5\% of detailed models from \cite{pols_1998}. This allows us to quickly generate large grids with thousands of tracks (each track takes approximately 5 seconds to complete) and determine a baseline requirement for a sufficient number of tracks for training and validating the emulator.

Below a mass of $0.8\ M_{\odot}$, the analytical tracks diverge from the behavior of stars at and above this mass due to different fitting relations, producing a discontinuity in the HR diagram. To avoid this artifact, we restrict the analytical grid to $0.8 \le M/M_{\odot} \le 1.2$. We also use their pre-provided system of EEP-like points for the full stellar evolutionary path, where a value of `1' represents the end of the MS. Using the analytical grid does not impact the development of an interpolator for stellar simulations, as the goal is to create a baseline to follow when using simulation-based grids.

Thus, our analytical grid consists of 2000 tracks with approximately 100 models per track, evenly spaced in mass, $M\in [0.8,1.2]\ M_{\odot}$, for a constant metallicity, $Z=0.02$. We only include up to the end of the main sequence (e.g., TAMS), and do not use mass loss, though it is possible to include it. The top left panel in Figure~\ref{fig:hr_other_grids} shows 50 representative tracks within this grid. To avoid introducing any additional uncertainty, we do not resample or interpolate \textit{along} the track (in stellar age, an age proxy, etc.). While this makes for an imperfect model-density comparison to the \texttt{MESA} grids, it is the best way to isolate the uncertainty injected by the interpolation methods themselves.

\subsubsection{MESA Grids}
\label{sec:mesa_grid}

We generate a 200-track grid using \texttt{MESA} 
\citep[Modules for Evolution in Stellar Astrophysics, version r24.03.1,][]{paxton_2011,Paxton2013,Paxton2015,Paxton2018, Paxton2019,Jermyn2023}. The \texttt{MESA} stopping condition is determined by the depletion of central hydrogen, defined by a central mass fraction of $X_c<10^{-3}$.  For our simple grid, \texttt{OPAL} opacities \citep{Iglesias_1993_opal,Iglesias_1996_opal} are used with $Z_{\rm init}=0.02$, $Y_{\rm init}=0.27$, $\alpha_{\rm MLT}=2.0$, and solar abundance mixture \citep{GS_1998G}. Overshoot is implemented as overmixing with an exponential profile \citep{Herwig_1999} core overshoot $(f_{\rm ov,core}=0.01)$, and shell overshoot $(f_{\rm ov,shell}=0.005)$. The outer boundary condition adopts the Eddington $T-\tau$ relation.

To ensure that the \texttt{MESA} tracks begin from the same evolutionary stage, we trim each track to span exactly from the zero-age main sequence (ZAMS) to the terminal-age main sequence (TAMS). ZAMS is identified using
\begin{equation}\label{eq:zams}
    X_c < 1- Z_0- Y_0 -0.01
\end{equation}
where $X_c$ is the central hydrogen abundance and $Z_0$ and $Y_0$ are the initial metallicity and initial helium abundance, respectively. As the exact model satisfying this condition differs per track, we use a local linear interpolation between the bracketing models to produce a consistent starting point. The stellar age is then shifted so that $t=0$ corresponds to ZAMS for every track. TAMS is identified analogously, when $X_c=10^{-3}$, and a similar interpolation ensures uniform termination points. On average, these tracks have 1,000 models (where the number of models per track ranges from approximately 800 to 1175). With 1,000 CPU cores and using a single CPU per track, a 1,000-track grid takes approximately 9 wall-clock hours and 6 GB of memory to run. These runtimes would vary depending on the available hardware.

To avoid introducing additional sources of uncertainty, we do not interpolate along the already dense \texttt{MESA} tracks. Thus, all interpolation burden falls on the cross-track direction, such that the ML methods described in Section~\ref{sec:interpolation} must learn to navigate between evolutionary tracks of neighboring masses, including through the non-monotonic transition region where the standard EEP construction fails. The smoothness and density of the grid in the mass direction, therefore, directly determine the difficulty of the interpolation problem, motivating the grid density analysis in Section~\ref{sec:grid_density}.

\subsection{Machine Learning Interpolation Methods}
\label{sec:interpolation}

Let the ground-truth simulator (e.g., from \texttt{MESA} or another generated grid) be defined as a mapping from the parameter input space to the observable output space 
\begin{equation}
    f: \Theta \to \mathcal{Y}, \quad \theta \mapsto f(\theta)
\end{equation}
where $\theta\in\Theta$ represents input parameters (e.g., mass, metallicity, etc.), and $y=f(\theta) \in \mathcal{Y}$ is the vector of output observables ($T_{\rm eff}, L, \Delta\nu,\nu_{\rm max}$). An emulator is a computationally efficient surrogate function, $\tilde{f}$, designed to approximate $f$ based on a finite training set $\mathcal{D}= \left\{ (\theta_i, y_i) \right\}_{i=1}^n$ computed at discrete points, $i$, within the grid. Because $\mathcal{D}$ is finite, $\tilde{f}$ must interpolate between grid points, which can introduce error. The severity of this error depends on the local structure of $f$: in regions where $f$ is highly non-linear, interpolation is less accurate. We therefore model the emulator output as
\begin{equation}
    \tilde{f} (\theta) =f(\theta) + \epsilon_{\rm em}(\theta)
\end{equation}
where $\epsilon_{\rm em}(\theta)$ represents the interpolation error or structural discrepancy. As we will show, some regions of parameter space are more challenging to interpolate than others due to degeneracies.

\begin{deluxetable*}{rccccccc}

\tabletypesize{\footnotesize}
\tablewidth{0pt}

 \tablecaption{Specifications for the machine learning models used within this work for grid interpolation.  \label{tab:ml_model_specs}}

\tablehead{
 \colhead{Hyperparameters} & \colhead{Linear} & \colhead{kNN} & \colhead{RF} & \colhead{FLAX A} & \colhead{FLAX B} & \colhead{FLAX C} & \colhead{FLAX D} }

 \startdata
 Neighbors & & 2 & & & & &\\
 Weights & & Distance & & & & &\\
 p & & 2 & & & & &\\
 \hline
 Estimators & & & 989 & & & &\\
 Max. depth & & & 16 & & & &\\
 Min. samples split & & & 3 & & & &\\
 Min. samples leaf & & & 2 & & & &\\
 Max. features & & & SQRT & & & &\\
 Max. samples & & & 0.7 & & & &\\
 \hline
 Batch size & & & & 128 & 64 & 256 & 64\\
 Hidden dims. & & & & (128,64,32,8) & (128,64,32,16,8) & (128,64,32,8) & (128,64,32,8)\\
 Learning rate (LR) & & & & \texttt{5e-4} & -- & \texttt{3e-3} & \texttt{5e-4} \\
 Scheduler & & & & cosine & -- & -- & -- \\
 Warm-up frac. & & & & 0.03 & -- & -- & -- \\
 End LR frac. & & & & 0.05 & -- & 0.3 & 0.05 \\
 Loss & & & & MSE & WMSE & MSE & --\\
 Epochs & & & & 10,000 & -- & -- & -- \\
 EMA decay & & & & 0.97 & -- & -- & -- \\
 Weight decay & & & & 0 & -- & -- & --\\
 Dropout & & & & 0 & -- & -- & -- \\
 Update WMSE weights per & & & & -- & 5 epochs & -- & --\\
 WMSE power & & & & -- & 2 & -- & --\\
 WMSE weight clip & & & & -- & (0.05, 20) & -- & --\\
 \enddata

 \vspace{0.2cm}
 \tablecomments{For FLAX, all models use the \texttt{gelu} activation function. WMSE refers to the weighted-MSE scheme, outlined in Section~\ref{sec:training_pipeline}. FLAX Version B shows the changes from Version A, and so on. The kNN and RF hyperparameter values were determined by \texttt{Optuna}, based on the 200-track \texttt{MESA} grid. Additionally, the FLAX models were run using \texttt{JAX v0.7.2} on one NVIDIA RTX 5000 Ada GPU on the Yale Bouchet HPC.}
 
\end{deluxetable*}
To evaluate the sensitivity of interpolation error to model complexity, we apply four methods of increasing flexibility: linear interpolation, k-nearest neighbors (kNN), random forests (RF), and \texttt{JAX}-based neural networks (FLAX). Each method and the relevant hyperparameters are summarized in plain terms in Appendix~\ref{sec:model_specs}.

\subsubsection{Interpolator Training Pipeline}
\label{sec:training_pipeline}

A key design choice in our interpolation pipeline is that we treat mass and age as simultaneous inputs to a single 2D interpolation, rather than adopting the two-step approach used in some prior works \citep{rodrigues_2017, Maltsev_2024, clara_2025}, which first interpolates along tracks (in age or an age proxy) and then across tracks (in mass). While this two-step approach is intuitively appealing as it respects the physical structure of the grid by treating the along-track and cross-track directions separately, it introduces additional complexity. Any errors introduced in the first interpolation step propagate into the second. 

By contrast, treating mass and age as joint inputs allows the ML methods to learn the full 2D mapping directly from the data, without imposing assumptions about the separability of the two directions. This becomes crucial in the transition region, where along-track and cross-track behaviors are strongly coupled. The age at which a star develops a convective core depends on its mass, and the resulting non-monotonicity in $X_c$ means that a single age value can correspond to qualitatively different evolutionary states depending on which side of the transition the stars fall on. We describe the shared pipeline in Appendix~\ref{sec:full_interp_pipeline}, which notes any method-specific differences where relevant.

\subsubsection{Interpolation Errors on the Test Set}
\label{sec:interpolation_errors_on_test_set}

Interpolation errors are quantified using absolute-valued fractional residuals between the emulator predictions and the underlying stellar model grid used for testing. These residuals are directly compared to the minimum observational uncertainties from the \textit{Kepler}-LEGACY asteroseismic sample of MS stars \citep{Lund_2017}, thereby placing interpolation errors in an astrophysically meaningful context. For $\Delta\nu$, the best targets from \textit{Kepler}-LEGACY have a fractional uncertainty on the order of 0.1\%. This comparison allows us to assess whether the numerical precision of the interpolated grid is sufficient for precision asteroseismology and its applications. 

If all of the per-model interpolation residuals of the test set are smaller than the minimum observational uncertainties, the grid can be considered adequately smooth and dense for precision inference. Conversely, if residuals exceed this observational error floor, interpolation error becomes a limiting factor and biases inferred stellar parameters. 

Beyond evaluating the overall magnitude of the residuals, we also examine their distribution within the input parameter (here, mass and age) space, and identify the location of the largest interpolation errors. Mapping residual structure across these dimensions allows us to diagnose regions of sparse sampling, rapid structural evolution, or sharp transitions in stellar properties that may degrade interpolation performance. Including other varying input parameters, like metallicity or mixing length, will likely further influence interpolation smoothness without additional refinement.

\subsection{Uncertainty}
\label{sec:uncertainty}

While residuals provide a diagnostic of the interpolator's local performance, they also form the basis for a formal treatment of the emulator errors inherent in the surrogate modeling process. To ensure robust stellar characterization, these approximate errors must be integrated into a comprehensive uncertainty framework alongside observational and physical constraints.

In this problem, there are four types of uncertainty ($\sigma$), all of which must be correctly quantified and propagated: observational uncertainty (data noise), emulator uncertainty (ML-method imperfection), physics uncertainty (assumptions embedded in the stellar grid itself), and inference uncertainty (posterior geometry in parameter space). Only the first two types enter the likelihood, but all four contribute to the structure and interpretation of the final posterior distributions. 

Observational uncertainty, $\sigma_{\rm obs}$, is the random and irreducible noise in the measured stellar quantities, or our output parameters ($T_{\rm eff}, L, \Delta \nu, \nu_{\rm max}$). We construct an observational covariance matrix
\begin{equation}
    \Sigma_{\rm obs} = {\rm diag}\left(\sigma_1^2 , \sigma_2^2,... \right)
\end{equation}
which is propagated directly into the likelihood. This matrix sets the baseline width of the posterior distribution and dominates the total uncertainty budget when emulator errors are negligible (which is the ideal scenario). In practice, if correlations between observed quantities are known, they can also be incorporated through non-zero off-diagonal terms.

Emulator uncertainty, $\sigma_{\rm em}$, arises from limited training data, finite model capacity, and stochastic optimization. The neural network approximates the mapping
\begin{equation}
    f(\theta) = \left(T_{\rm eff}, L, \Delta \nu, \nu_{\rm max} \right)
\end{equation}
where $\theta$ is the set of input parameters (e.g., mass, age, etc.), so that the emulator prediction satisfies
\begin{equation}
    y_{\rm NN} (\theta) = y_{\rm true} (\theta) + \epsilon_{\rm em}(\theta)
\end{equation}
where $\epsilon_{\rm em}(\theta)$ represents the approximation error at a single point $\theta$ for a single trained network. This true error $\epsilon_{\rm em}(\theta)$ is only directly measurable at grid points in the test set; we characterize it through the residual distribution $\hat{y}_{\rm NN}(\theta_i)-y_i$ on these test-set points, which sets a floor on the approximation error and reflects how it varies across parameter space. In addition to this approximation error, we capture uncertainty due to stochastic variation in training using an ensemble of $N$ independently trained networks differing only in random seed. The predictive covariance across the ensemble,
\begin{equation}
    \Sigma_{\rm em} = {\rm Cov}(y_{\rm NN})
\end{equation}
quantifies sensitivity to training stochasticity, capturing variance and cross-output correlations induced by the model. We note that $\Sigma_{\rm em}$ and the test set residual distribution are complementary: the former reflects stochastic uncertainty, while the latter reflects systematic approximation error. If $\Sigma_{\rm em}$ does not adequately capture the residual distribution, particularly in regions of high non-linearity, the uncertainty model may be misstated. The total output covariance entering the likelihood is therefore
\begin{equation}\label{eq:total_output_covariance}
    \Sigma_{\rm total} = \Sigma_{\rm obs} + \Sigma_{\rm em}
\end{equation}
and defines a single linearized consensus posterior. We also evaluate the ensemble mixture distribution (in the case of neural networks), defined as the average of the individual posteriors from each network,
\begin{equation}
    P(\theta \mid D) = \frac{1}{N}\sum_{i=1}^N \mathcal{N}(\theta \mid \hat{\theta_i}, \Sigma_{i})
\end{equation}
where $\theta$ is the parameter vector (independent variable of the PDF), $\hat{\theta_i}$ is the maximum a posteriori (MAP) estimate, or mean parameter vector produced by the $i$-th neural network, and $\Sigma_i$ is the posterior covariance associated with the $i$-th network. For additional detail about computing the likelihood for an ensemble mixture model, see Appendix~\ref{sec:uncertainty_mixture_models}.

\section{Results}
\label{sec:results}

We visualize the interpolation residuals using the Cumulative Distribution Function (CDF), where we refer to the \textit{``CDF value"} as the value that denotes the fraction of test-set models with absolute-valued fractional residuals below the minimum observational uncertainty. 

We report the fractional residual as a signal-to-noise ratio (SNR) relative to the minimum fractional \textit{Kepler}-LEGACY uncertainty ($\sigma_{\rm obs}/\sigma_{\rm em}$). An SNR $\gg1$ indicates that the emulator error is negligible and is safe for high-precision missions. An SNR $\approx 1$ indicates that the emulator is on-par with the best data, though caution is needed in hierarchical fits. Finally, an SNR < 1 indicates a dominating emulator error and results biased by interpolation artifacts.

We discuss the importance of grid density in Section~\ref{sec:grid_density}, uncertainty quantification and propagation in Section~\ref{sec:uq}, and the interpolation of pre-computed grids in Section~\ref{sec:other_grids}. All results until Section~\ref{sec:other_grids} will use the various subsets of the same 2000-track analytical grid, and the same set of 399 test tracks from this grid will be used for consistency.

\subsection{Quantifying the Success of ML Methods}
\label{sec:quantify_ml_method}

To quantify the success of various machine learning methods, we define our metric as the number of predictions with absolute fractional residual values that are lower than the best fractional uncertainty in $\Delta \nu$ from \textit{Kepler}-LEGACY, on the order of 0.1\%.

\begin{figure}
    \centering
    \includegraphics[width=\linewidth]{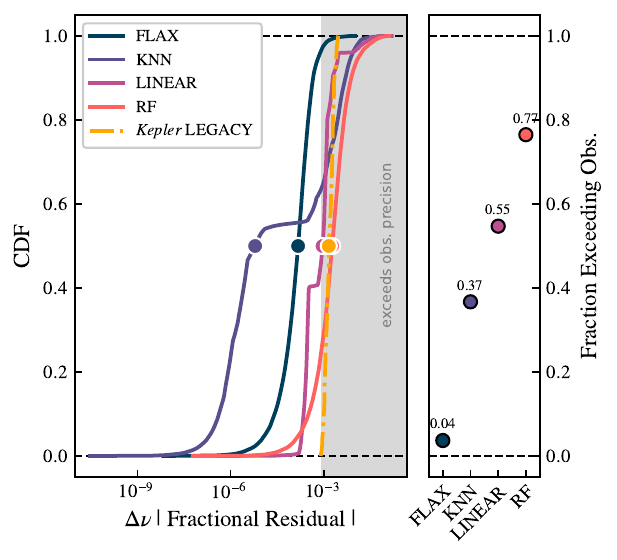}
    \caption{Cumulative Distribution Function (CDF) for the absolute-valued fractional residuals from four ML-methods: linear interpolation, kNN, RF, and FLAX A. Each was trained on a 200-track subset of the full 2000-track analytical grid. \textit{Left:} Each curve shows the CDF for the $\Delta\nu$ absolute value of fractional residuals for the test data. The shaded region delineates the values of these residuals that exceed the observational precision. The dot on each curve represents the median residual value for each method. \textit{Right:} A summary plot where each point is labeled with the fraction of models in the test set with residuals that exceed the observational precision.}
    \label{fig:hurley_compare_ml}
\end{figure}

\begin{figure*}
    \centering
    \includegraphics[width=0.99\linewidth]{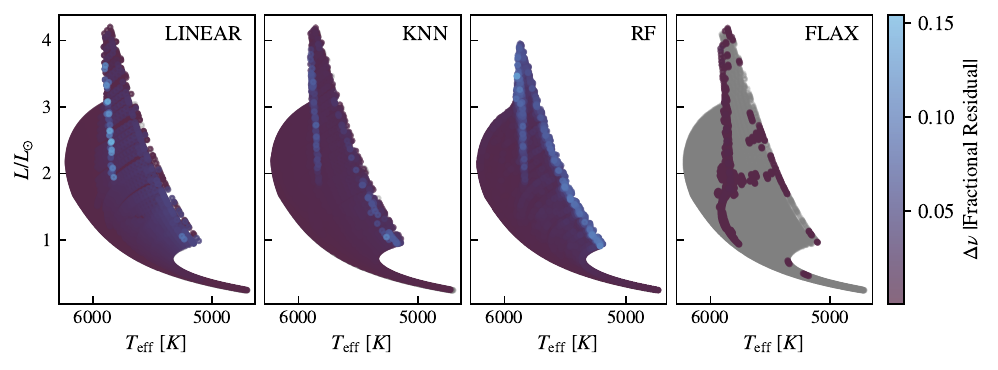}
    \caption{HR diagram of interpolated test set tracks, as obtained by linear interpolation, kNN, RF, and FLAX A, colored by the absolute-valued fractional residual of $\Delta\nu$. A subset of 200 tracks from the 2000-track analytical grid is used for training and validation. The gray points, mainly visible with FLAX, show the models within each test set track that have an absolute-valued fractional residual below the minimum fractional $\Delta\nu$ observational uncertainty from \textit{Kepler}-LEGACY.}
    \label{fig:analytical_hr_multi_ML}
\end{figure*}

For a preliminary test, we use 200 of the remaining tracks after setting aside the consistent testing set, for training and validation (161 and 39 tracks, respectively). We use linear interpolation, kNN, RF, and FLAX A (see Table~\ref{tab:ml_model_specs} for specifications) as interpolators, such that they can be used to predict $T_{\rm eff}$, $L$, $\Delta \nu$, and $\nu_{\rm max}$ of the test set. The main panel of Figure~\ref{fig:hurley_compare_ml} shows these results as a CDF plot, spanning from 0 to 1. The best-case scenario is that the entirety of any CDF curve should be to the left of the shaded region that denotes residuals exceeding observational precision. 

However, as none of these curves fall into the best-case scenario, the location at which they cross the observational precision line becomes important. This provides a broad understanding of the percentage of models within the test set that have an absolute-valued fractional residuals that are greater than the observational precision. The $\Delta\nu$ CDF value for the linear interpolation line is 0.452, indicating that approximately 45\% of models within the testing set have absolute-valued fractional residuals that are less than the minimum fractional $\Delta\nu$ \textit{Kepler}-LEGACY uncertainty. The $\Delta\nu$ CDF values are 63\% for kNN, 24\% for RF, and 96\% for FLAX. The right panel offers a complementary summary to the CDFs, where a point is plotted per method to show the fraction of models within the test set that have residuals exceeding the observational uncertainty.

Linear interpolation, kNN, and RF all had SNR values that indicate a dominant emulator error: 0.005, 0.007, and 0.006, respectively. FLAX has an SNR of 0.07, which is a significant improvement over the other machine learning methods, but would still yield results that are emulator-error dominated and biased by interpolation artifacts. 

Determining how emulator residuals compare with observational uncertainties provides a baseline for emulator performance. However, the location of these high-residual models within the larger grid is more important in obtaining a deeper understanding of how to construct more informative grids and more flexible emulators for precise asteroseismology. Figure~\ref{fig:analytical_hr_multi_ML} shows an HR diagram of the testing set colored by the emulator's absolute-valued fractional residuals in $\Delta\nu$ for linear interpolation, kNN, RF, and FLAX. The location of the highest-residual models, along the transition between radiative and convective cores, is most apparent for the FLAX results, where most of the test set has residuals lower than the minimum \textit{Kepler}-LEGACY uncertainty.

\begin{figure}
    \centering
    \includegraphics[width=\linewidth]{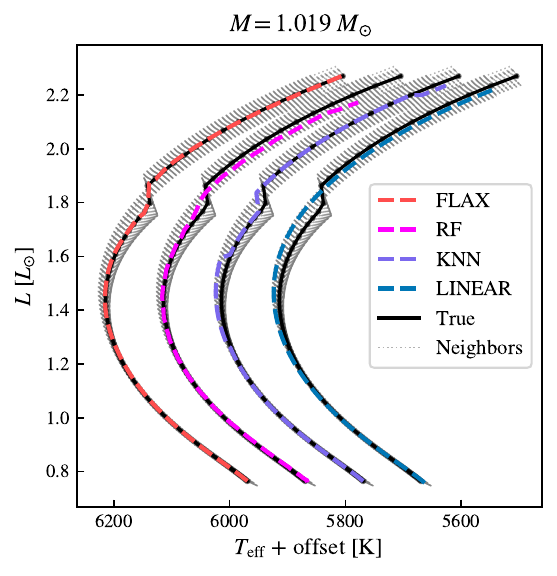}
    \caption{Comparison of the same true test track of mass $M=1.0189\ M_{\odot}$ to the predicted tracks from linear interpolation, kNN, RF, and FLAX. For readability, we offset each set of true + predicted tracks for each of the ML methods. A subset of 200 tracks from the 2000-track analytical grid is used for training and validation. The gray, dashed lines are the 50 training tracks that surround this test track. }
    \label{fig:predicted_vs_true_track}
\end{figure}

Figure~\ref{fig:predicted_vs_true_track} shows the interpolation ability of each ML method for a test track of mass $M=1.019\ M_{\odot}$, in the transition region between having a radiative core (smooth and continuous increase in luminosity) and having a convective core (discontinuous jump in luminosity). The panels also include 10 tracks from the training set that surround this test track. Linear interpolation begins relatively well until the luminosity reaches 1.25 $L_{\odot}$. Beyond this point, linear interpolation predicts that this track has a radiative core. Additionally, it does not fully reach the TAMS location of the true track and instead stops short. The kNN method shows improvement but exhibits a bifurcation in the predicted path, reflecting the instability of the nearest-neighbor selection when the test point lies between disparate evolutionary regimes. Like linear interpolation, the kNN-interpolated track truncates prematurely relative to the TAMS. The RF model avoids the bifurcation seen in kNN by consistently mapping the star to a convective core regime, yet it deviates from the true track right before and after the discontinuous jump in luminosity. FLAX, while not perfect, shows a significant improvement in all regions where the other methods struggled, including reaching the correct TAMS location.

\subsection{Grid Density}
\label{sec:grid_density}

To determine how many analytical tracks are necessary to achieve test-set residuals below the observational uncertainty, we train FLAX networks using the restricted training sets of 25, 200, and 500 tracks as defined in Section~\ref{sec:training_pipeline}, increasing the number of tracks available for training and validation while holding the test set fixed. 

\begin{figure}
    \centering
    \includegraphics[width=\linewidth]{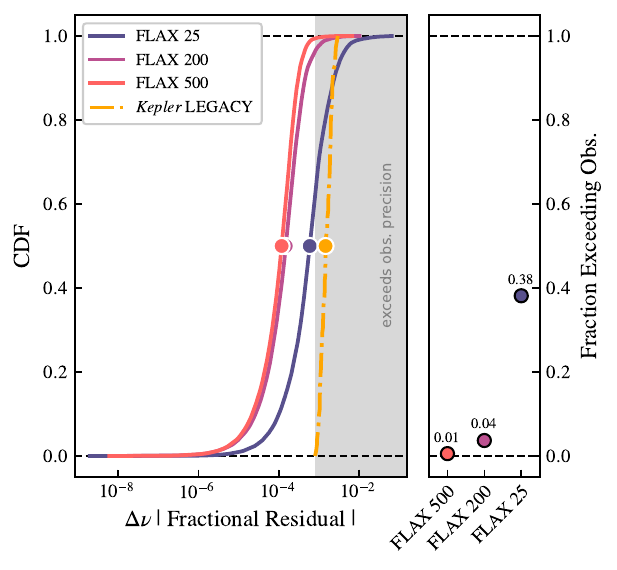}
    \caption{CDF for the absolute-valued fractional residuals of $\Delta\nu$ as we increase the size of the available training and validation sets. From the 2000-track analytical grid, FLAX 25 (network C, 25 tracks), FLAX 200 (network A, 200 tracks), and FLAX 500 (network B, 500 tracks) were each tested using the same test set. A summary plot is included, showing the fraction of test-set models with residuals exceeding the observational uncertainty: 1\%, 4\%, and 38\% for FLAX 500, 200, and 25, respectively.}
    \label{fig:cdf_hurley_compare_train_size}
\end{figure}

Figure~\ref{fig:cdf_hurley_compare_train_size} shows the CDF of the absolute-valued fractional residuals as we increase the size of the available training and validation data sets. FLAX 25 (network C) has 25 tracks (21 for training and 4 for validation), FLAX 200 (network A) has 200 tracks (161 for training and 39 for validation), and FLAX 500 (network B) has 500 tracks (401 for training and 99 for validation). To see the full configuration details, see Table~\ref{tab:ml_model_specs}. The $\Delta\nu$ CDF value for FLAX 25 is 0.618, indicating that approximately 62\% of models within the testing set have absolute-valued fractional residuals that are less than the minimum fractional $\Delta\nu$ \textit{Kepler}-LEGACY uncertainty. For FLAX 200, this value increases to 96\%, and for FLAX 500, we see that 99\% of testing set models have residuals lower than the minimum fractional \textit{Kepler}-LEGACY uncertainty. 

There is a similar improvement in the median fractional residual. We obtain an SNR of 1.35 for FLAX 25, indicating that with 25 tracks used for training and validation, the emulator error is comparable to the observational error and thus needs to be used with caution for hierarchical fits. For FLAX 200, the SNR increases to 5.55, and for FLAX 500, the SNR further increases to 7.1. Increasing the size of the training and validation set results in emulators with errors that are of the same order of magnitude as observational errors, but are not entirely immune to future high-precision missions (e.g. \textit{PLATO}) where there will be many more observations of stars in the transition region.

We also note the SNR for comparing the minimum \textit{Kepler}-LEGACY uncertainties to the maximum fractional residuals. For FLAX 25, the SNR is 0.01, indicating that the worst-case interpolation suffers significantly from dominant emulator error. For FLAX 200, the SNR is increased to 0.07, which is still well within the range for dominant emulator error. For FLAX 500, the SNR is again increased to 0.1. Although this is not a sufficiently high SNR for the emulator errors to be considered on-par with the observational errors, it indicates that uniformly increasing the grid density does not entirely remove the ridge of models in the transition region that have residuals higher than the observational uncertainty.

The utility of an emulator is defined by its computational efficiency relative to direct simulation. If we let $N_{\rm stars}$ be the number of targets to be characterized and $N_{\rm eval}$ be the number of model evaluations required per star (e.g., within an MCMC sampler), the cost of an ``on-the-fly" approach is $N_{\rm stars} \times N_{\rm eval}$. For the emulator to be a viable alternative, the number of simulations required to train the grid, $N_{\rm grid}$, must satisfy $N_{\rm grid} \ll N_{\rm stars} \times N_{\rm eval}$. If $N_{\rm grid}$ approaches or exceeds this product, the overhead of grid generation defeats the purpose of the surrogate model. Consequently, rather than simply increasing the global density of the training set, we discuss more sophisticated strategies for handling these ``difficult" regions in Section~\ref{sec:improvements}.

While the global metrics discussed in this section demonstrate the overall ability of various ML methods in emulating, or interpolating, stellar tracks, the picture of a method's utility for scientific inference is incomplete. A method may achieve a low mean squared error across a broad training set, yet still exhibit localized instabilities that only become apparent when the emulator is used to infer stellar parameters from observations. In Section~\ref{sec:uq}, we therefore shift focus from residual statistics to inference reliability, quantifying how emulator uncertainty propagates into the posterior distribution on stellar mass.

\subsection{Uncertainty Quantification and Propagation}
\label{sec:uq}

Uncertainty quantification requires a distinction between random uncertainty (the fixed noise in seismic observations) and emulator uncertainty (the ML-model disagreement caused by gaps in the training data). Mass inference requires inverting the ML-model to find the input mass that matches a specific seismic observation. Therefore, the reliability of the result is tied to the local slope, or Jacobian, of the interpolation. In regions of sparse training data, NNs introduce small-scale numerical instabilities that are almost invisible when plotting the predicted track with the true track. However, these instabilities are magnified during the inference process: they cause fluctuations of the local gradient, blowing up small observational errors into large mass uncertainties, and creating false solutions. \textbf{Additional details regarding mass inference can be found in Appendix~\ref{sec:uncertainty_mixture_models}.}

\begin{figure}
    \centering
    \includegraphics[width=\linewidth]{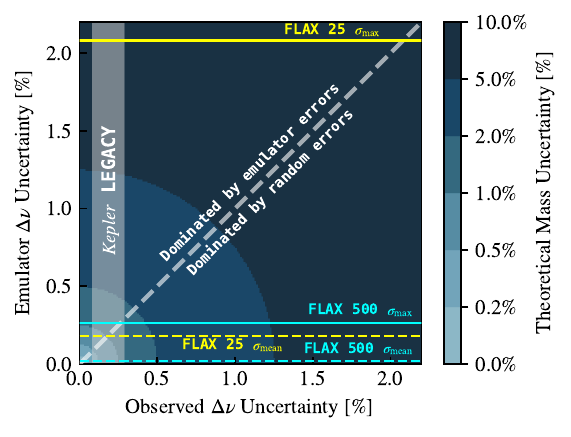}
    \caption{The scaling mass uncertainty for the 2000-track analytical grid as a function of observational and emulator $\Delta\nu$ uncertainties. Above (below) the white, the dashed line represents mass uncertainties dominated by emulator (random/observational) uncertainty. We plot the maximum (solid) and mean (dashed) ensemble model uncertainties for FLAX 25 (yellow) and FLAX 500 (cyan). We show the range of \textit{Kepler}-LEGACY uncertainties as the gray shaded region.}
    \label{fig:2D_uncertainty}
\end{figure}

To provide an intuition for how the $\Delta\nu$ uncertainty propagates, Figure~\ref{fig:2D_uncertainty} shows a simplified mapping of the expected theoretical mass uncertainty as 
\begin{equation}
    \sigma_M/M\approx 4\sqrt{\sigma_{\Delta\nu,{\rm obs}}^2 + \sigma_{\Delta\nu, {\rm em}}^2}
\end{equation}
for any given observational and emulator $\Delta\nu$ uncertainties. It is important to note that this plot serves as a sensitivity analysis rather than a full statistical characterization. To isolate the effect of $\Delta\nu$, we assume all other output parameters ($T_{\rm eff}$, $L$, and $\nu_{\rm max}$) are known without error. Furthermore, while the standard scaling relation between $\Delta\nu$ and mass (Eq.~\ref{eq:Delta_nu}) is analytical and thus technically does not include any emulator uncertainty, we use its sensitivity ($|\partial \ln(M)/\partial\ln(\Delta\nu)|\approx 4$) as a proxy to visualize how the emulator's internal scatter propagates into a final mass estimate. 

The diagonal represents an equal contribution from both components, while above (below) the diagonal represents masses dominated by emulator (random/observational) errors. Thus, the ideal scenario for any given interpolation is that the total mass uncertainty is dominated by random errors from observation, and not from the emulator itself. 

\subsubsection{FLAX Ensemble Models}
\label{sec:flax_ensemble_models}

\begin{figure*}
    \centering
    \includegraphics[width=\linewidth]{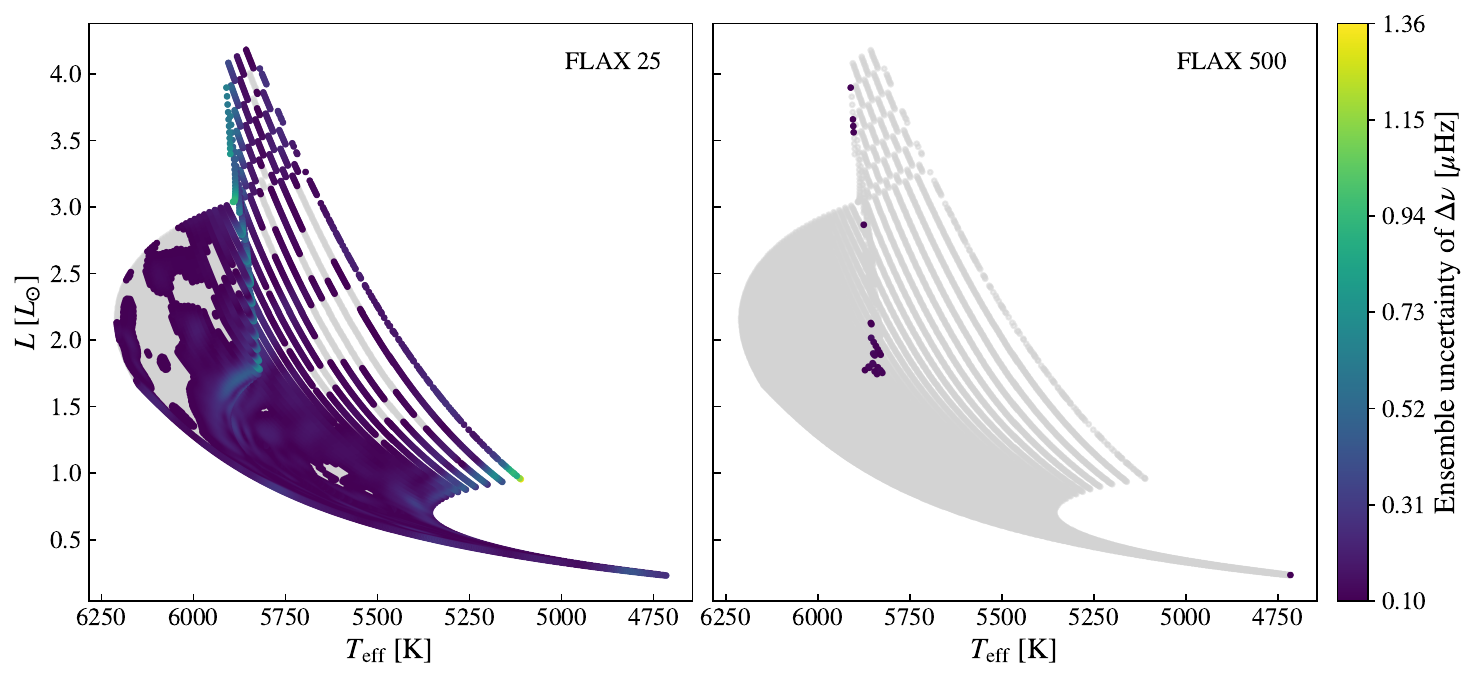}
    \caption{HR diagrams of the 399-track analytical grid test set for ensembles of FLAX 25 (left) and FLAX 500 (right), colored by the ensemble uncertainty of $\Delta\nu$. We set an ensemble uncertainty floor of 0.1 $\mu{\rm Hz}$, such that models with an uncertainty below it are instead plotted in gray.
    }
    \label{fig:analytical_hr_ensemble_std}
\end{figure*}
We create two ensembles of 15 FLAX neural networks each, only varying the random seed that controls the training process. The number of ensemble members is determined by calculating the per-test-track residual divided by the ensemble uncertainty as a function of increasing the number of included members. The standard deviation of this quantity approaches 1.0 if the ensemble accurately estimates the emulator uncertainty. With 15 members, the aforementioned quantity is 1.1 for $\Delta\nu$, indicating that the ensemble moderately underestimates its emulator uncertainty. Although going beyond 15 members does continue to reduce this quantity closer to 1.0, there are diminishing returns with each additional ensemble member. The longer training time and computational cost is not worth the marginal improvement in uncertainty calibration. 

One ensemble is based on the FLAX 25 configuration, and the other is based on the FLAX 500 configuration. They are tested on the same test set of 399 tracks. The goal is to quantify the impact of a lack of training data across the parameter space. Fig.~\ref{fig:2D_uncertainty} overplots the maximum and mean uncertainties obtained by the two emulators. It is clear that while the means for FLAX 25 and FLAX 500 are similar, the maximum uncertainties are not. For the \textit{Kepler}-LEGACY uncertainty in $\Delta\nu$ (shaded band), characterizing a star's mass using FLAX 500 results in a mass uncertainty of up to 1-2\%, while the mass uncertainty using FLAX 25 is up to 10\%.

Figure~\ref{fig:analytical_hr_ensemble_std} shows an HR diagram to map the uncertainty level of the predicted $\Delta\nu$ from the test set within each ensemble. The left panel shows the FLAX 25 ensemble, where a ridge of higher ensemble uncertainty appears. It starts at the base of the transition region and continues upwards at the location where stars with convective cores have a discontinuous jump in luminosity. This is the same region that had higher absolute-valued fractional residuals in Fig.~\ref{fig:analytical_hr_multi_ML}, most visible with the FLAX model. It is no coincidence that the region with the highest residuals overlaps with the region that shows the most disagreement between ensemble model members. The right panel shows the FLAX 500 ensemble, where the ridge seen with FLAX 25 has essentially disappeared. With FLAX 500, there are only a handful of individual models in which the ensemble uncertainty was above the 0.1 $\mu{\rm Hz}$ floor. These few models still lie within the transition region of radiative-to-convective cores.

\begin{figure*}
    \centering
    \includegraphics[width=0.99\linewidth]{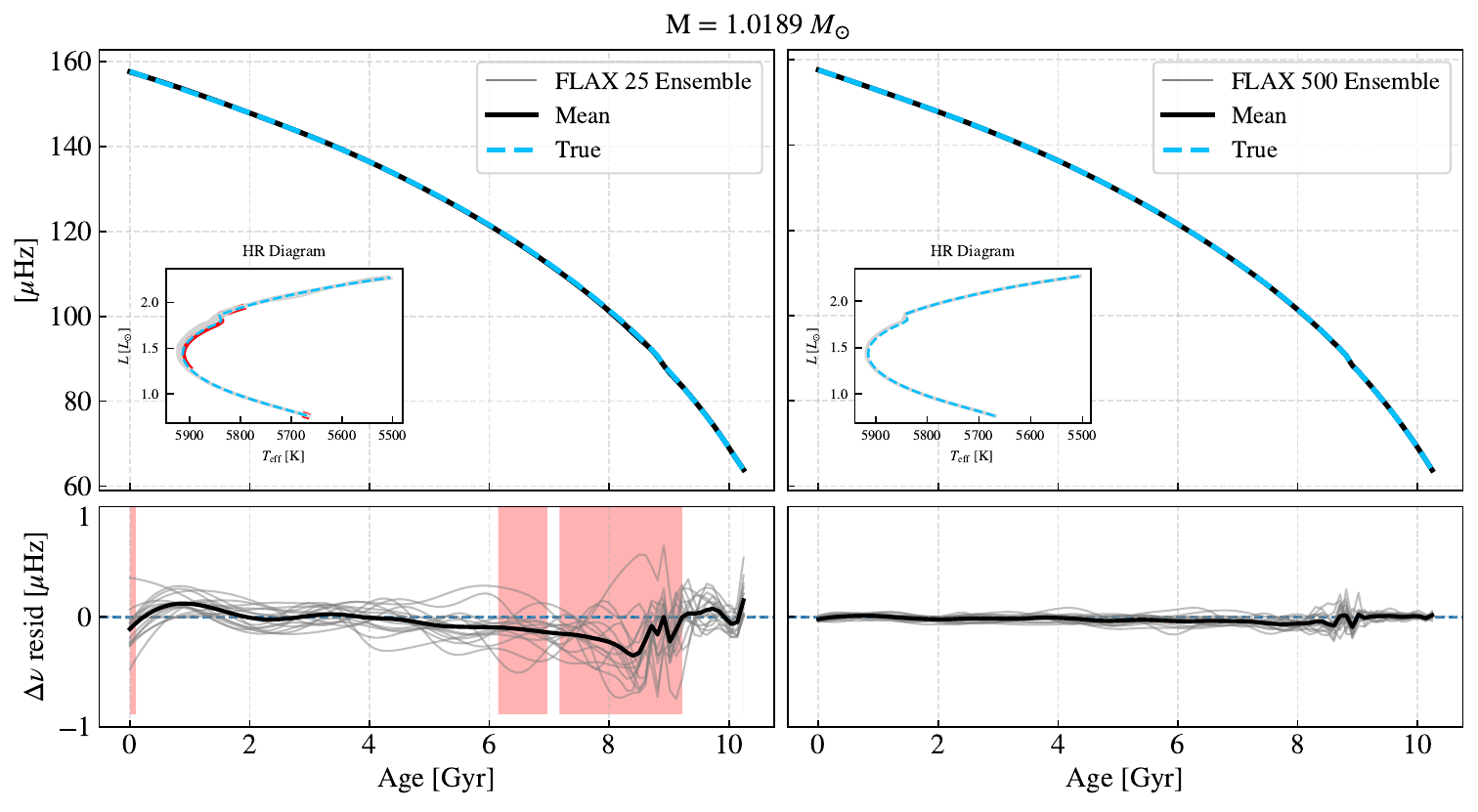}
    \caption{Comparison of ensemble model results FLAX 25 (left) and FLAX 500 (right) for a $M=1.0189\ M_{\odot}$ star in the analytical grid test set. Top panels show the true, ensemble mean, and individual ensemble predictions of $\Delta \nu$ as a function of star age. Bottom panels show the residuals (individual and ensemble member values minus the true value) for the individual ensemble members. The red shaded region on the lower left residual panel highlights where the residuals in $\Delta \nu$ are larger than half of the maximum residual value across the ensemble members. That same region is plotted in red within the HR diagram.}
    \label{fig:compare_track_1095_ensemble}
\end{figure*}

Figure~\ref{fig:compare_track_1095_ensemble} shows a direct comparison of the two ensembles as they perform on a track within the transition region, corresponding to a star of $1.0189\ M_{\odot}$. The top left panel of Fig.~\ref{fig:compare_track_1095_ensemble} shows the results for the FLAX 25 ensemble. The residuals in the lower left panel range from approximately $[-1,1]$, where residuals are simply calculated by subtracting the individual ensemble model member predictions (or the ensemble mean) from the true value of $\Delta \nu$. We highlight the ages within the evolution that have residuals higher than a residual threshold (here, $=\pm0.48$), calculated as half of the maximum residual value over all ensemble members. We plot these ages in the inset HR diagram in red as well, to show that this corresponds to the portion of the evolutionary track that is different between stars with a purely radiative core and those with a convective core. Stars with a purely radiative core will not have the sharp increase in luminosity. 

The right panel of Fig.~\ref{fig:compare_track_1095_ensemble} shows the improvement in the same track when using FLAX 500. There is no shaded region in the residual panel, as none of the residuals in this ensemble went beyond the threshold value calculated by the FLAX 25 ensemble. We see a tighter agreement between the individual members of the ensemble model, with significant improvements in the region where FLAX 25 struggled.

\subsubsection{Posterior Distributions and Model Consensus}
\label{sec:post_distributions}

\begin{figure*}
    \centering
    \includegraphics[width=\linewidth]{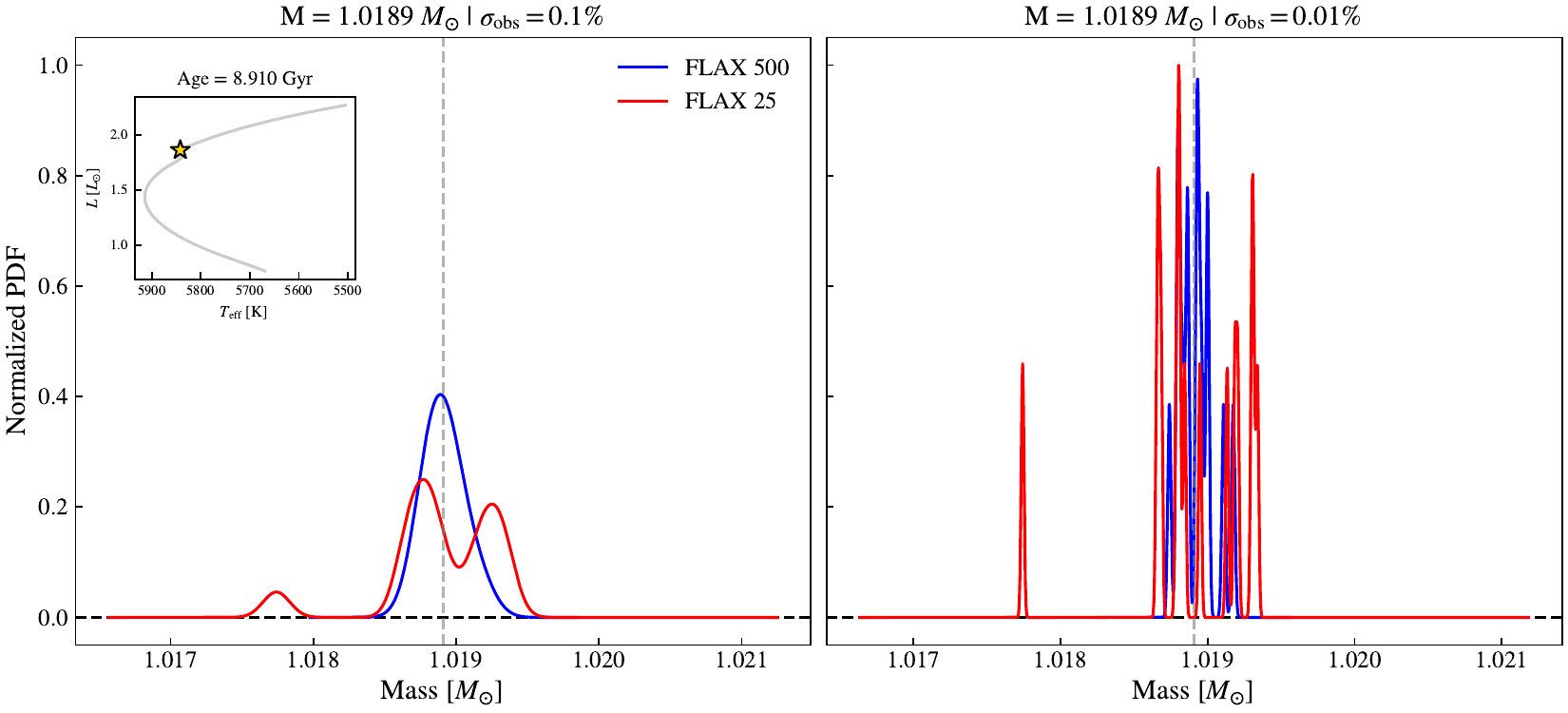}
    \caption{Comparison of normalized mass posterior distributions using only the $\Delta \nu$ contribution for an analytical grid test-set star ($M=1.0189\ M_{\odot}$, age $= 8.910\ {\rm Gyr}$; see inset HR diagram) using the FLAX 25 and FLAX 500 ensembles. \textit{Left}: Distributions under a 0.1\% observational uncertainty for $\Delta \nu$. \textit{Right}: Reducing the observational uncertainty by an order of magnitude (0.01\%) exposes discretization artifacts in the underlying grid. The dashed vertical line indicates the true stellar mass. This shows that grid density becomes important as one decreases the uncertainties.}
    \label{fig:track_1095_post_dist}
\end{figure*}

Figure~\ref{fig:track_1095_post_dist} shows how the differences in predicted $\Delta\nu$ from each ensemble member of FLAX 25 and FLAX 500, seen in Fig.~\ref{fig:compare_track_1095_ensemble}, translate into probability distributions of the inferred stellar mass at a particular stellar age. The posterior distribution function (PDF) for each ensemble is computed as a mixture of Gaussians, where each ensemble member contributes an individual Gaussian centered at the mass inferred from its predicted $\Delta\nu$ and the local Jacobian. We normalize each PDF to integrate to 1.0, and then divide by the global maximum across both panels such that the PDF ranges from [0,1]. Since the global maximum is set by the sharpest posterior across both panels, the broader posteriors in the left panel do not peak at 1.0. Given the relative heights, a taller peak indicates stronger ensemble consensus, and a peak aligned with the ``true mass" (the mass as given in the testing set) indicates an unbiased inference. Multiple peaks indicate disagreement among the ensemble members, with each peak corresponding to a distinct cluster of predicted $\Delta\nu$ values.

The left side of Figure~\ref{fig:track_1095_post_dist} compares the normalized PDFs for a star of true mass $M=1.0189\ M_{\odot}$ and age $8.91\ {\rm Gyr}$, from two ensemble models. To isolate the effect of uncertainty in $\Delta\nu$ on the mass, the Jacobian ignores the other output parameters. The observational parameter values for this star used to characterize its mass are: $T_{\rm eff} = 5841.82\ {\rm K}$, $L=1.862\ L_{\odot}$, $\Delta\nu = 88.399\ \mu{\rm Hz}$, and $\nu_{\rm max} = 872.342\ \mu{\rm Hz}$. The inset HR diagram shows the true mass and age of the inferred star being characterized. 

For the FLAX 500 ensemble, the posterior is unimodal and peaks at approximately $1.0189\ M_{\odot}$. This single peak indicates a strong consensus among the ensemble members. In contrast, the FLAX 25 ensemble produces a wide, multi-modal distribution with three peaks. This is a direct result of the instability within the Jacobian, which created multiple local regions where the predicted $\Delta \nu$ matches the observation, leading the model to suggest several contradictory mass solutions. These results demonstrate that while sparse data may produce a visually plausible track, it lacks the derivative stability required for precision asteroseismic inference.

We perform a sensitivity analysis by reducing the observational uncertainty to 0.01\%, an order of magnitude less than \textit{Kepler}-LEGACY, such that the main source of error must come from the neural network itself. The right side of Fig.~\ref{fig:track_1095_post_dist} shows the results of this experiment. FLAX 500 now has multiple peaks ranging between 1.0187 and 1.0192 $M_{\odot}$ where the highest central peak remains at 1.0189 $M_{\odot}$. This behavior indicates that the ensemble has largely reached a consensus where the error budget is no longer dominated by observational (random) uncertainty, but is instead approaching the noise floor of the interpolation. FLAX 25 has 15 peaks (corresponding to the seven ensemble members) ranging between 1.0177 and 1.0193 $M_{\odot}$, where each peak is much sharper. This behavior confirms that for low-density training sets, the error budget is entirely dominated by emulator uncertainty. Specifically, it creates interpolation artifacts where the emulator's manifold lacks the resolution to track the likelihood surface. These results suggest that for future high-precision missions (e.g., PLATO), the density of the training grid is as critical as the choice of the physical model itself, as observational gains can be entirely dominated by emulator-induced biases.

It should be noted that although the spread of the ensemble member predictions is relatively contained, even for FLAX 25 (spanning approximately 1.018--1.02 $M_{\odot}$), this performance represents a ``best-case" scenario for a sparse ensemble. Additionally, this may represent an artificial precision floor created by the lack of competing physical solutions in a simplified grid. In this sparse regime, the FLAX 25 emulator is unable to capture the nuanced non-linearities of stellar evolution, particularly in the transition region.

While Fig.~\ref{fig:compare_track_1095_ensemble} highlights how increasing the data available for training and validation from 25 to 500 tracks resolves discrepancies between ensemble model members, the broader impact of this is best understood through its effect on the final inferred stellar parameters. The ensemble spread defines an intrinsic emulator uncertainty for a single model evaluation, which must be below the observational uncertainty for precision stellar characterization. Regardless of how small the observational uncertainty is, a dominant emulator uncertainty will lead to an incorrect characterization. 

Having established the grid density requirements for reliable emulation and uncertainty propagation using controlled analytical grids, we now ask whether the widely-used precomputed grids available to the community -- \texttt{ASTEC}, \texttt{MIST}, and \texttt{YREC} -- meet these requirements for precision asteroseismology in the transition mass range.

\subsection{Interpolating Precomputed Grids}
\label{sec:other_grids}

We now show the results of interpolating other widely-used grids: \texttt{ASTEC}, \texttt{MIST}, and \texttt{YREC} (see Appendix~\ref{sec:model_specs} and Fig.~\ref{fig:hr_other_grids} for details). 
\begin{figure}
    \centering
    \includegraphics[width=\linewidth]{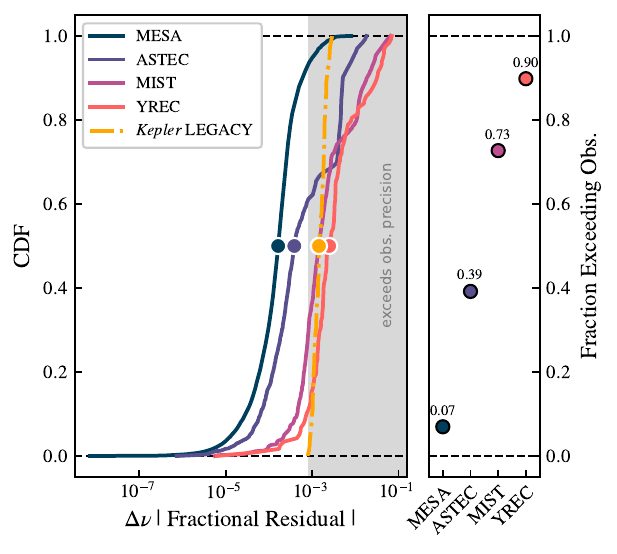}
    \caption{\textit{Left:} The CDF for the $\Delta\nu$ absolute-valued fractional residuals for the 2D cases of the \texttt{ASTEC}, \texttt{MIST}, \texttt{YREC}, and \texttt{MESA} (see Section~\ref{sec:mesa_grid}) grids. We used interpolator configurations FLAX A, FLAX D, FLAX D, and FLAX C, respectively (see Table~\ref{tab:ml_model_specs} for details). \textit{Right:} A summary plot, as in Fig.~\ref{fig:hurley_compare_ml}.}
    \label{fig:cdf_flax_compare_other_grids}
\end{figure}
Figure~\ref{fig:cdf_flax_compare_other_grids} compares the CDF for the $\Delta\nu$ absolute-valued fractional residuals for the 2D cases of the \texttt{ASTEC}, \texttt{MIST}, \texttt{YREC}, and \texttt{MESA} grids, as described above. The \texttt{MESA} grid generation is discussed in Section~\ref{sec:mesa_grid}.

These results demonstrate that the grid density requirements are not met in the transition mass range under \textit{Kepler}-LEGACY precision requirements. This has direct implications for any stellar characterization that relies on these grids without accounting for interpolation uncertainty. In Section~\ref{sec:discussion}, we contextualize these findings within the broader landscape of prior interpolation efforts, and discuss strategies for addressing the grid density problem in future work.

\section{Discussion}
\label{sec:discussion}

\subsection{Interpolation Efforts}
\label{sec:previous_works}

Previous interpolation efforts span a wide range of algorithmic choices, grid configurations, and evaluation criteria, making direct comparison challenging. Nonetheless, we compare our results quantitatively, especially noting the precision metric applied, and whether emulator uncertainty is propagated into the final stellar characterization.

\cite{clara_2025} is the most directly comparable work, as their \texttt{MESA}-generated grid has a narrow mass range ($1.15-1.35\ M_{\odot}$), which includes part of our transition region defined by the analytical grid, at fixed metallicity. Their best configuration (cubic spline along tracks, linear interpolation across tracks, with central density as the age proxy) returns a maximum absolute frequency error of $1.2\ \mu{\rm Hz}$. Additionally, they use the residual as the sole diagnostic for their mass posterior. In our work, the FLAX 25 ensemble achieves uncertainties in $\Delta\nu$ well below $1\ \mu{\rm Hz}$ for most of the test set, and FLAX 500 reduces this further (see Fig.~\ref{fig:analytical_hr_ensemble_std}). We also explicitly propagate the interpolation error into the mass posterior, in addition to the test-set residuals.

\cite{rodrigues_2017} extends to three input dimensions (mass, age, and metallicity), complicating direct comparison to our work. Their grid, which consists of only 21 evolutionary tracks per metallicity, results in a 5\% mass precision and 19\% age precision. Bringing this 3D grid into the 2D case of only varying initial mass and age, it becomes comparable to our FLAX 25 scenario where mass uncertainties are approximated to be at the 5-10\% level (see Fig.~\ref{fig:2D_uncertainty}).

Instead of emulating the forward model, \cite{bellinger_2016} uses random forests in the inverse direction to learn the mapping from observables directly to stellar parameters in a 6D grid. This RF inverse mapping produces a point estimate; thus, the uncertainty cannot be propagated into the posterior. Our comparison of RF performance in Section~\ref{sec:quantify_ml_method} (CDF score 0.24 for $\Delta\nu$) confirms that RF is less accurate than FLAX as a \textit{forward} emulator in the transition region. However, the 6D grid is substantially richer in physical realism, and the direct comparison in Section~\ref{sec:other_grids} (which uses the 20-track \texttt{YREC} subset for a $\Delta\nu$ CDF value of 0.102) illustrates the compounded challenge of interpolating over a sparse grid with a less flexible model.

In the likely case where degeneracies exist within the grid, such that different combinations of input parameters are able to produce the same observable quantity, machine learning methods that offer one-to-one predictions, like random forest, are not sufficient. \cite{Li_2022} explicitly test their uncertainty calibration using Gaussian process (GP) regression and find that it severely underestimates uncertainties in the 3D (mass, EEP, $Z_0$) and 5D (mass, EEP, $Z_0$, $Y_0$, and $\alpha_{\rm MLT}$) cases. Like \cite{clara_2025}, they also only rely on the test set errors as an estimate of the systematic uncertainty. On the other hand, our neural network ensemble approach computes calibrated, localized uncertainty estimates that correctly identify the transition region as the high-uncertainty zone (Fig.~\ref{fig:analytical_hr_ensemble_std}).

\cite{lyttle_2021} used a 4D (mass, $Z_0$, $Y_0$, and $\alpha_{\rm MLT}$) \texttt{MESA}-generated grid of 17,220 tracks to train an artificial neural network. Using the Sun as a test case, they achieve precisions of 2.5\% in mass, 1.2\% in radius, and 12\% in age. While they report uncertainties that appear well-calibrated, they do not explicitly propagate emulator error into the posterior covariance. \cite{scutt_2026} uses a branching architecture to separate classical outputs from asteroseismic (individual mode frequencies) components. They further reduce the individual mode frequencies by Principal Component Analysis \citep[PCA,][]{brown_1994} for training. However, they find that individual mode frequency uncertainties (0.3--1.1 $\mu{\rm Hz}$) are emulator-error dominated. This is exactly the failure mode our ensemble analysis identifies. Critically, neither work tests what happens when grid density is reduced below the level needed for the transition region. Our Figs.~\ref{fig:compare_track_1095_ensemble} and \ref{fig:track_1095_post_dist} provide the first explicit demonstration of how emulator-induced multimodality propagates into the mass posterior.

\cite{hon_2024} uses conditional normalizing flows with a 6D (mass, $Z_0$, $Y_0$, $\alpha_{\rm MLT}$, $f_{\rm ov, shell}$, and $f_{\rm ov, core}$) \texttt{MESA}-generated grid to capture correlations in high-dimensional output space for red giants. Their emulator errors are below 1\% in well-sampled regions, increasing near low-density or overlapping regions. While these results are quantitatively consistent with our findings, they do not test the specific radiative-to-convective core transition or provide a grid-density ablation. \cite{stone-martinez_2025} uses observational data instead of a simulated \texttt{MESA}-generated grid. This choice avoids the high computational costs of generating a sufficiently dense grid, but requires thousands of well-characterized stars. The upper range of their age uncertainties (18-30\%) corresponds to sparse regions of parameter space, justifying our pipeline for in-depth analyses of grid-density requirements.

\cite{Maltsev_2024} identify the same mass transition we study, noting that $M_{\rm init}/M_{\odot} \in (1.16, 1.5)$ requires a finer grid spacing than the rest of the \texttt{MIST}-based grid. To mitigate this, they generate additional tracks in this region for a $\delta M=0.01\ M_{\odot}$. However, their evaluation metric is the percent-precision on their output parameters ($L$, $T_{\rm eff}$, and $\log(g)$), and does not reference observational uncertainty thresholds. This makes the adequacy of their added tracks ambiguous for the specific science goal of asteroseismic characterization at \textit{Kepler}-LEGACY precision. \cite{kallinger_2026} ran a similar nested-sampling algorithm to focus on high likelihood regions of parameter space. Their Bayesian inference pipeline was aimed at understanding the uncertainties stemming from near-surface effects, eliminating the need for ad-hoc surface corrections, thus improving asteroseismic characterization.

Across the aforementioned prior works, the interpolation errors are reported as residual statistics without a specific observational precision threshold. Additionally, they are generally not propagated in a way directly comparable to the posterior-level emulator-error treatment used here. We address the lack of specific observational context by adopting the minimum fractional \textit{Kepler}-LEGACY observational uncertainty as an explicit threshold for interpolation residuals. This enables us to assess the emulator performance directly within the context of an observational survey rather than reporting errors in isolation. We then provide a systematic grid-density ablation that identifies both the magnitude of interpolation failure and its location within the parameter space. From this, we show that the transition region between radiative and convective cores is the dominant source of emulator error across all methods and grid configurations tested. Finally, and most consequentially for future high-precision missions, we propagate emulator uncertainty into stellar mass posteriors, revealing that emulator-induced multimodality is a qualitatively different and more dangerous failure mode than elevated mean residuals, one that cannot be diagnosed from residual statistics alone.

\subsection{Improvements in Track Selection During Training}
\label{sec:improvements}

The goal of grid interpolation is to reduce the number of tracks needed in the first place. However, currently, any interpolation scheme requires a full grid of tracks to interpolate between, and generally building a grid implies a set spacing between tracks, whether in stellar mass, $Z_0$, $\alpha_{\rm MLT}$, etc. This set spacing, especially when considering the two-dimensional case of only varying initial mass and age, is unnecessary, as half of the grid is filled with ``easy" tracks (low mass, radiative core). Under these conditions, when selecting tracks for training, validation, and testing, the emulator wastes precious computational resources fitting the numerous easy tracks.

In this work, tracks are generated based on an even spacing in stellar mass. Alternatively, tracks can be generated using a Sobol \citep{Sobol_1967} sequence such that input parameter values are chosen by a quasi-random distribution of points \citep{bellinger_2016, BASTA_2022}. However, not all regions of the parameter space are equally challenging to interpolate, and as previously indicated, there is a more methodical way to generate tracks. Tracks with fully radiative cores (below the transition region to convective cores, where the threshold is dependent on many other properties) follow a monotonic pattern on the MS. The transition region between radiative and convective cores is much less trivial, as there may be degeneracies depending on the physics used. Thus, fewer low-mass tracks should be needed to train the model, and a high density of tracks should be generated on-the-fly, as needed, in the transition region, when the emulator calculates high residuals using the validation set.

We aim to incorporate this scheme in future iterations of this work, as its extension to a higher-dimensional input grid is crucial to avoid generating millions of tracks. A version of such active learning \citep[see][for review]{settles_2009} has been implemented within the context of stellar simulations, and is effective in improving efficiency and lowering computational costs \citep{Rocha_2022}. Grid generation and training will always take some amount of time; adaptive sampling seems to reduce this time, and this is a one-time cost that then enables ``free" inference. Incorporating such adaptive strategies would directly address the bottleneck identified in Section~\ref{sec:grid_density}, ensuring that computational resources are concentrated precisely where the interpolation problem is most difficult.

\section{Conclusions}
\label{sec:conclusions}

In this work, we have quantified the requirements for accurate and precise emulators for interpolating asteroseismic parameters, contextualized by observational uncertainties. We focus on the mass range $M\in[0.7,1.2]\ M_{\odot}$ because this range encapsulates where the onset of convective cores produces non-monotonic evolutionary behavior seen in the central hydrogen fraction, $X_c$ (see Fig.~\ref{fig:hr_tracks_Xc_abundance}). Linear interpolation, k-Nearest Neighbors (kNN), random forest regression (RF), and FLAX neural networks are used to interpolate 2D (only varying the initial stellar mass and age) grids generated by \texttt{MESA}, an analytical grid from \cite{Hurley_2000}, as well as grids from \texttt{YREC}, \texttt{ASTEC}, and \texttt{MIST}. The grids include input variables mass and age (linear in Gyr), and output variables $T_{\rm eff}$, $L$, $\Delta\nu$, and $\nu_{\rm max}$. Our main results in this work using the analytical grid are the following:
\begin{enumerate}
    \item There is a clear advantage in using FLAX neural networks when interpolating analytical stellar evolution grids for precision asteroseismology. With a 200-track training set, they achieve a much higher $\Delta\nu$ CDF score (96\%) than linear interpolation (45\%), kNN (63\%), and RF (24\%) (Fig.~\ref{fig:hurley_compare_ml}).
    \item The models with the highest residuals lie along the transition region between stars born with radiative cores and those with convective cores (Fig.~\ref{fig:analytical_hr_multi_ML}).
    \item The size of the training set severely impacts the performance of neural networks in interpolating across tracks. The $\Delta\nu$ CDF value increases from 62\% to 99\% as the training set size increases from 25 to 500 tracks (Fig.~\ref{fig:cdf_hurley_compare_train_size}).
    \item Uncertainties are propagated to obtain posterior distributions of mass via ensemble models of neural networks. Given a sparse grid of training data, the propagated uncertainty from $\Delta\nu$, composed of the \textit{Kepler}-LEGACY uncertainty and the emulator uncertainty, yields multi-modal peaks in the mass posterior for a star in the transition region. Increasing the size of the training data resolves the multimodality (Fig.~\ref{fig:track_1095_post_dist}).
\end{enumerate}

A natural extension of this work is the emulation of individual mode frequencies, rather than the global asteroseismic parameters $\Delta\nu$ and $\nu_{\rm max}$, used here. High-level asteroseismic fitting typically involves direct comparison to individual frequencies, which carry substantially more information about stellar interior structure than the global scaling relations alone. Emulating individual frequencies introduces additional challenges beyond those demonstrated here: the number of output dimensions grows with the number of observed modes, the frequencies are more sensitive to fine structure details near the core transition region, and the mapping from stellar parameters to individual frequencies is more nonlinear than the scaling-relation-based mapping used in this work. 

The ensemble uncertainty framework developed here, specifically the propagation of emulator covariance through the Jacobian into the stellar parameter posterior, is easily adaptable to this higher-dimensional output space, though the density requirements identified in Section~\ref{sec:grid_density} are likely to be more stringent. The recent work of \cite{scutt_2026}, who emulate individual radial modes using a branching neural network and find emulator-dominated uncertainties at high radial order, suggests that grid density in the transition region will be an even more critical bottleneck for individual frequency emulation than it is for the global parameters studied here.

We also emphasize that the challenges demonstrated here represent the most favorable interpolation scenario: a two-dimensional input space varying only stellar mass and age, limited to the main sequence, with all other physical parameters held fixed. Varying other input dimensions like metallicity, helium abundance, mixing length, and convective overshoot will introduce new regions of rapid structural change and additional degeneracies. The grid density requirements identified here are therefore a lower bound on what will be needed for a production emulator. If emulator-induced multimodality and unstable Jacobians arise in this simplified setting, one should expect them to be more severe, more widespread, and harder to diagnose as the dimensionality of the input space grows.

These considerations point to a broader lesson for the field: the reliability of an emulator cannot be assessed independently of the science case it is intended to serve. The adequacy of grid density, algorithm choice, and uncertainty quantification is defined by the precision demands of the observational survey at hand. As remarkably precise asteroseismic surveys grow in size, the emulators underpinning stellar characterization must be held to correspondingly higher standards. Explicit precision thresholds, localized uncertainty diagnostics, and ensemble-based posterior propagation must be integral to any interpolation or emulation pipeline. The transition region between radiative and convective cores studied here is governed by rapidly changing physics. Understanding and resolving the interpolation failure in this region is a prerequisite for the precise and unbiased stellar characterization that future surveys will demand.

\section*{Acknowledgments}
We thank J{\o}rgen Christensen-Dalsgaard for valuable comments and feedback. This work was supported by a Yale AI Seed Grant (Project No. PJ117090) and the Yale Center for Research Computing. This material is based upon work supported by the U.S. National Science Foundation under award No.\ 2606207.

%%%%%%%%%%%%%%%%%%%%%%%%%%%%%%%%%%%%%%%%%%%%%%%%%%
%\section*{Data Availability}

\appendix
\section{Model Specifications}
\label{sec:model_specs}

This appendix provides brief, self-contained descriptions of the four interpolation methods used in this work, along with definitions of the relevant hyperparameters, intended for readers less familiar with machine learning.

\begin{enumerate}
    \item \textit{Linear Interpolation} \citep{Virtanen_2020_scipy}: Given a set of data points with known output values, linear interpolation estimates the output at a new location by determining nearby data points by Delaunay triangulation \citep{bardford_QUICKHULL_1996}. It then computes a proportionally weighted average of their values based on how close the new point is to the surrounding points. Unlike machine learning methods, there is no model ``training" in the traditional sense, and nothing is ``learned". The data points directly define the interpolant without any optimization or hyperparameter tuning. A held-out test set is still used to evaluate the performance. It is fast and exact at the data points, but struggles significantly when the underlying relationship is nonlinear. 

    \item \textit{k-Nearest Neighbors \cite[kNN,][]{Altman_kNN_1992}}: kNN predicts the output at a new point by averaging the outputs of the $k$ closest points in the training set, where closeness is measured by Euclidean distance in input space. ``Training" here consists of storing the training data, and during prediction, the algorithm searches for the $k$ nearest neighbors. The $k$ parameter for the number of nearest neighbors is tunable and controls how local or global the prediction is. Since closeness is measured by Euclidean distance in input space, the inputs must be scaled, especially if they span a wide range of magnitudes. Without scaling, finding a nearest neighbor becomes an impossible task. Note that training time scales linearly with sample size as $\mathcal{O}(N)$. Hyperparameters include:
    \begin{enumerate}
        \item The number of neighbors, $k$: controls how local or global the prediction is, where smaller values produce more local but potentially noisier predictions, while larger values produce smoother but potentially oversmoothed predictions. 
        \item Weights: The ``distance" weighting scheme determines how the contributions of $k$ neighbors are combined, such that closer neighbors receive proportionally higher weight.
        \item $p$: Controls the Minkowski distance metric. $p=2$ is the standard Euclidean distance.
    \end{enumerate}

    \item \textit{Random Forests \citep[RF,][]{{biau_RF_2012}}}: A random forest is an ensemble of decision trees, where each tree acts like a flowchart. Each tree partitions the input space into rectangular regions by learning a sequence of threshold rules (e.g., ``if mass $< 1.01\ M_{\odot}$, go right"), and predicts the average output within each region. Individual trees tend to overfit, so averaging the predictions of many trees trained on random subsets of the data and input features helps to reduce variance and improve generalization. Here, training scales with sample size on the order of $\mathcal{O}(N\log N)$, and consists of constructing these trees. Prediction is simply a lookup through the learned rules. No input scaling is required here, since the tree splits depend on the ordering of values, and not their magnitudes. This can easily be implemented in Python using \texttt{sklearn} \citep{Pedregosa_2011_sklearn}. Hyperparameters include:
    \begin{enumerate}
        \item Number of estimators: Controls how many trees are grown.
        \item Maximum depth: Limits the number of splits each tree can make. Deeper trees fit the training data more closely, but are also more prone to overfitting.
        \item Minimum samples split/leaf: Minimum number of training examples required to split a node/leaf. Both parameters prevent the tree from making splits based on too few data points.
        \item Maximum features: Controls how many input features are considered at each split, where the square root of the total number of features (SQRT) is used.
        \item Maximum samples: Specifies the fraction of training data used to grow each tree.
    \end{enumerate}

    \item \textit{Neural Networks with JAX (FLAX)}: \texttt{JAX} \citep{bradbury_jax_2018} is an open-source Python library developed for high-performance numerical computation. It provides automatic differentiation (computing the gradients of arbitrary functions automatically), and compiles numerical code to run efficiently on GPUs and TPUs. \texttt{FLAX} \citep{flax2020github} is a neural network library built on top of \texttt{JAX} that provides modular building blocks (layers, activations, optimizers) for defining and training neural networks. 

    A neural network maps inputs to outputs through a sequence of \textit{layers}. Each layer computes a weighted sum of its inputs and passes the result through a nonlinear \textit{activation function}, where the weights are the learnable parameters of the model. We contrast this with linear interpolation, where the weights are limited to linear functions. Training consists of adjusting these weights to minimize a loss function (e.g., MSE between predicted and true outputs for regression tasks) using gradient descent. The gradient of the loss with respect to all weights is computed automatically by \texttt{JAX}, and the weights are updated iteratively throughout the training process. The depth (number of layers) and width (number of units per layer) of the network control its learning capacity and are tuned (either through manual tuning or via grid-search algorithms like \texttt{Optuna} \citep{akiba_2019_optuna}) as hyperparameters. 

    Neural networks are the most flexible of the four methods, and can capture complex nonlinear relationships, while efficiently scaling to large datasets on GPU hardware. However, they require more careful tuning and longer training times than the other methods described here. Additional training details, including the weighted loss function, exponential moving average of weights, and hyperparameter selection, are described in Section~\ref{sec:training_pipeline}. Hyperparameters include:
    \begin{enumerate}
        \item Batch size: The number of training examples processed in each gradient update step. Smaller batches introduce more stochastic noise into the gradient estimates to escape local minima, while larger batches produce more stable but potentially less exploratory updates.
        \item Hidden dimensions: The depth (number of layers) and width (number of units per layer) of the network, together controlling the learning capacity.
        \item Learning rate (LR): Controls the step size of each gradient update.
        \item LR scheduler: Instead of a constant LR throughout the entire neural network training, a scheduler is used to vary the LR. We use cosine decay to gradually reduce the LR following the cosine curve, preventing the optimizer from making large destabilizing updates when the network is closer to convergence \citep{Loshchilov_2016}.
        \item Warm-up fraction: The proportion of total training steps during which the learning increases linearly from zero to its initial value before decay begins. This stabilizes early training by preventing large gradient updates before the network weights have reached a reasonable initialization.
        \item End LR fraction: The minimum LR as a fraction of the initial value at the end of the cosine decay schedule, such that training continues to make small refinements in later epochs.
        \item Loss function: The objective minimized during training. The difficulty-weighted variant of the MSE is described in Section~\ref{sec:training_pipeline}.
        \item Epochs: Sets the number of complete passes of the entire dataset to train and validate the ML model, updating the weights after each epoch to improve accuracy.
        \item EMA decay parameter: Controls the exponential moving average of network weights \citep{Polyak_1992}. Values closer to 1 retain more history and produce smoother weight trajectories. Applying EMA during evaluation and inference reduces the impact of small fluctuations in stochastic gradient descent updates, improving the generalization to unseen data.
        \item Weight decay: A regularization term that penalizes large network weights. We set this to 0, as otherwise the network does not converge.
        \item Dropout: A regularization technique that randomly sets a fraction of network activations to zero during training. Similarly, it is also set to 0.
        \item Update WMSE weights per: Determines how frequently the loss function weights are updated. Increased frequency can improve accuracy, but with a high computational cost.
        \item WMSE power: Controls how aggressively difficult examples are up-weighted relative to the mean residual, with higher values concentrating gradient updates on the most poorly predicted regions of parameter space. This is especially important as the transition region between radiative and convective cores is narrow with respect to the rest of the parameter space.
        \item WMSE weight clip: Prevents any single example from dominating the gradient update, preserving stable optimization dynamics throughout training \citep{Zhang_2020}.
    \end{enumerate}
\end{enumerate}

\section{Full Interpolation Pipeline}
\label{sec:full_interp_pipeline}

The full grid is split by track, ensuring that each evolutionary track appears entirely in either the training, validation, or testing set. This choice reflects the physical task of interpolating between tracks and prevents information leakage across evolutionary stages. Edge tracks (the first and last tracks of the grid if tracks are sorted by mass) are always included in the training set to avoid extrapolation during validation or testing. To ensure that we have the same tracks within the testing set for all analyses, we set aside $20\%$ of the full grid. The rest of the data is used for training ($64\%$) and validating ($16\%$) the model. 

In some analyses, we further restrict the training and validation set to a subset of tracks drawn at equal mass intervals across the full grid, referred to as a restricted training set. This allows us to systematically study the effect of grid density on interpolation performance while holding the test set fixed (20\% of the full grid). The restricted training sets used in Section~\ref{sec:grid_density} contain 25, 200, and 500 tracks, respectively, corresponding to average mass spacings of 0.02, 0.002, and 0.001 $M_{\odot}$. The validation set in each case is composed of 20\% of the restricted training set.

Scalers are fit only on the training set after the training, validation, and testing data have been established, preventing data leakage. The fitted scaler is then applied to the full grid before training any interpolator. The scaler type, either robust or standard, is chosen based on the distribution and skewness of any given parameter. Note that for tree-based methods (e.g., RF), scaling the data isn't necessary, as tree splits depend on order, not scale. For distance-based methods like kNN, scaling is necessary as the distance magnitude depends on scale. Similarly, scaling is also necessary for gradient and linear models like neural networks and linear or logistic regression, as gradients and regularization depend on scale.

For FLAX neural network models, there are many hyperparameters that can be tuned to optimize the training stability, convergence speed, and generalization performance. These include learning rate and its scheduling (e.g., constant, cosine decay, or warm-up schedules), weight decay for L2 regularization, dropout rate, network architecture parameters (number of hidden layers, hidden units per layer, activation functions), batch size, and optimizer configuration \citep[e.g., AdamW momentum parameters,][]{loshchilov_2019_adamw}. Additional information on the selected architectures and hyperparameter values for the best-performing networks can be found in Table~\ref{tab:ml_model_specs}.

Hyperparameters for kNN and RF are tuned via cross-validation. While FLAX hyperparameters (learning rate, weight decay, dropout, architecture, depth/width, etc.) can be optimized using automated hyperparameter search frameworks like \texttt{Optuna} \citep{akiba_2019_optuna}, we found that manual tuning was often more efficient in practice given the relatively well-behaved loss landscape. The training time is proportional to both the number of tracks and the number of models per track. 

We train FLAX by minimizing a weighted regression loss in scaled output space. For each training example $i$ with residual vector $\mathbf{r}_i = \hat{\mathbf{y}}_i - \mathbf{y}_i$, the base loss is computed per example by averaging the chosen per-output mean squared error loss across the output dimensions. To emphasize difficult training examples, we introduce a per-example weight $w_i$ derived from current model residuals. Periodically, a scalar difficulty metric (e.g., mean or percentile absolute residual) is computed per example on the validation set. Difficulties are scaled relative to the median residual and transformed via a power law $w_i \propto d_i^{\alpha}$, then clipped to a bound interval to ensure numerical stability and renormalized to have unit mean so that the effective learning rate is preserved. The final objective minimized at each step is the weighted average loss
\begin{equation}
    \mathcal{L} = \frac{1}{N}\sum_{i=1}^N w_i \ell_i
\end{equation}
where $\ell_i$ is the per-example loss. This approach increases gradient contributions from systematically mispredicted regions of parameter space while maintaining stable optimization dynamics through clipping and normalization. For efficiency, all outputs are predicted jointly; we verified that joint prediction does not degrade performance relative to single-output models for the cases examined here.

Note that training the analytical grids requires approximately 10 minutes to 75 minutes, on 10 CPUs, depending on the size of the training set. Training the \texttt{MESA}-\textbf{based interpolator, where the grid has}, more models per track \textbf{than the analytical grid}, takes significantly more time: 2-3 hours on a single GPU, or 8-9 hours on 10 CPUs. Testing with any grid is almost instantaneous. It should be noted that it is not possible to run FLAX on GPUs such that the results are exactly reproducible \citep{Shanmugavelu_arxiv_2024}. For fully reproducible results, one must use CPUs, but sacrifice computation time as the grid size increases.

\section{Other Grids}

\begin{figure*}
    \centering
    \includegraphics[width=\linewidth]{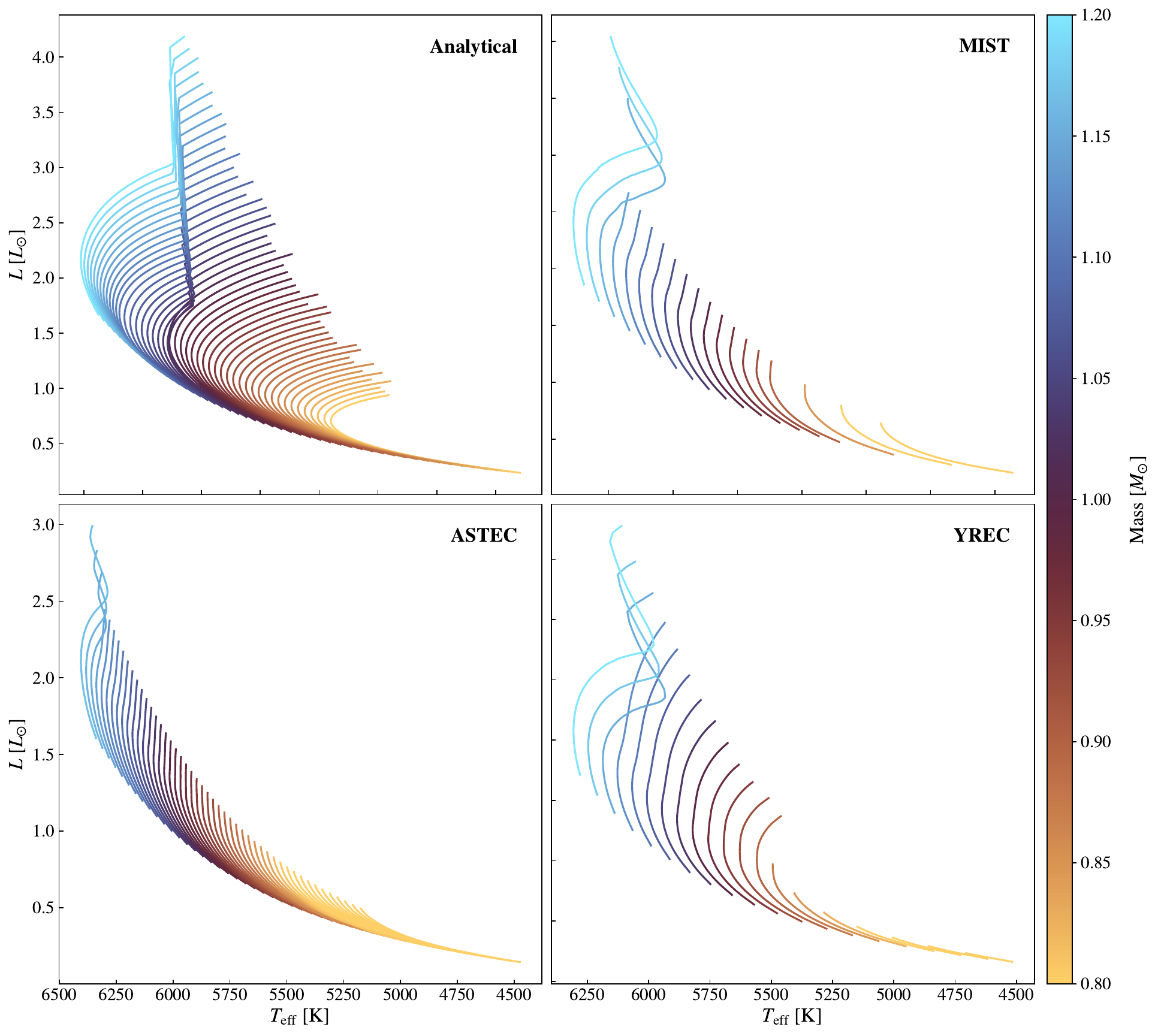}
    \caption{An HR diagram for the 2000-track analytical grid (top left) following \cite{Hurley_2000}, only showing 50 representative tracks for readability, the full 2D subsample of the \texttt{MIST} grid (top right), the full 2D subsample of the \texttt{ASTEC} grid (bottom left), and the full 2D subsample of the \texttt{YREC} grid (bottom right). The tracks are colored by mass.}
    \label{fig:hr_other_grids}
\end{figure*}

Figure~\ref{fig:hr_other_grids} shows HR diagrams for a subsection of the 2000-track analytical grid (50 tracks for readability), and all tracks available in the 2D cases of \texttt{MIST} (19 tracks), \texttt{ASTEC} (50 tracks), and \texttt{YREC} (20 tracks), colored by mass. We trained FLAX networks A, D, D, and C (see Table~\ref{tab:ml_model_specs}), respectively, with a data split between training, validation, and testing as follows: \texttt{ASTEC} (32,7,9), \texttt{MIST} (14,2,3), \texttt{YREC} (14,3,3), and \texttt{MESA} (126, 31, 38). The $\Delta\nu$ CDF scores are 0.608 for \texttt{ASTEC}, 0.272 for \texttt{MIST}, 0.102 for \texttt{YREC}, and 0.93 for \texttt{MESA}.

The \texttt{ASTEC} grid \citep{Christensen_Dalsgaard_2007} has 606 tracks in total, with $M\in [0.7,1.7]\ M_{\odot}$, $X=[0.694, 0.6978, 0.7017, 0.7056, 0.7094, 0.7133]$, $Z=0.0193$, and $\alpha_{\rm MLT}=2.1$. Bringing this into the 2D case and matching the \texttt{MESA} and analytical grids in their parameter ranges such that $M\in[0.7,1.2]\ M_{\odot}$ and $X=0.7017$, we are left with 48 tracks. For the training set, the average mass spacing in this grid subset is 0.01516 $M_{\odot}$, with a maximum spacing of 0.04 $M_{\odot}$. Although including tracks of varying $X$ in the same mass range would increase the grid density and the number of tracks we can allocate for training, varying $X$ results in additional ``difficult" regions. Thus, to study how well an interpolator can handle the transition region between radiative and convective cores, there must be no other additional degeneracies present.

From the \texttt{MIST} grid \citep{choi_2016}, out of 196 tracks with $Y=0.2703$, $Z=0.014286$, and $\alpha_{\rm MLT}=1.82$ to obtain the 2D case, we select the 19 available tracks of $M\in[0.7,1.2]\ M_{\odot}$. We only select phases 0-4 to evolve stars until TAMS. Note that \texttt{MIST} includes mass loss, so we note the initial mass spacing in the training set is 0.15 $M_{\odot}$. Since $X_c$ is available within the data, we use the same $X_c$-based TAMS cut as in Equation~\ref{eq:zams}.

From the \texttt{YREC} \citep{Demarque_2008} grid \citep[see Sec.~3.2 of][for details of the grid]{Metcalfe_2021}, out of 92 total tracks with $[{\rm Fe/H}]=0$ and solar $\alpha_{\rm MLT}$, we select the 20 available tracks with $M\in[0.8,1.2]\ M_{\odot}$. As this grid does not include $X_c$ as a variable, we estimate the TAMS location by a power-law, $L = C\ (T_{\rm eff})^k$, where $C$ is a non-zero constant. This grid has an average mass spacing for the training set of 0.036 $M_{\odot}$, with a maximum spacing of 0.075 $M_{\odot}$.

\section{Uncertainty of Ensemble Mixture Models}
\label{sec:uncertainty_mixture_models}

Comparing the consensus to the mixture allows us to diagnose cases where the networks disagree fundamentally (resulting in multi-modal or highly skewed distributions) rather than just exhibiting high variance. This provides a more robust check on the reliability of the emulator when using it to characterize stars. For the consensus case, the resulting likelihood surface (or posterior distribution) is then evaluated as:
\begin{equation}
    \mathcal{L}\propto {\rm exp}\left(-\frac{1}{2} \left( y^{\rm obs}- y^{\rm model}\right)^T \Sigma_{\rm total}^{-1} \left( y^{\rm obs}- y^{\rm model}\right) \right)
\end{equation}
Importantly, $\Sigma_{\rm em}$ is the \textit{correlated} emulator uncertainty, as it generally contains off-diagonal terms. For example, $\Delta \nu$ and $\nu_{\rm max}$ are correlated through stellar structure scaling relations, as are $T_{\rm eff}$ and $L$. These correlations manifest as tilted degeneracy directions in input parameter space and must be accounted for to avoid biased inference.

To determine if the total uncertainty in the output parameters, $\Sigma_{\rm tot}$, will significantly impact the uncertainty of the input parameters $\theta$ (e.g., mass, age, etc.), we propagate the output-space uncertainty through the Jacobian,
\begin{equation}
    \Sigma_{\theta} \approx \left( J^T \Sigma_y^{-1} J\right)^{-1}
\end{equation}
where $J=\partial y/\partial \theta$ and $\Sigma_y=\Sigma_{\rm total}$. This linear approximation corresponds to a local Gaussian (Fisher matrix) expansion of the posterior. This approach enables efficient comparison of emulator-induced uncertainty across multiple machine-learning models while preserving the local curvature structure of the likelihood surface. In the limit of a perfectly behaved emulator and small uncertainties, the linearized propagation and the ensemble mixture should converge. Discrepancies between these two methods serve as a diagnostic for non-linearities in the emulator's mapping or significant outliers within the ensemble.

Alternatively, $\Sigma_{\theta}$ can be obtained through full Bayesian sampling (e.g., MCMC), which captures non-linear degeneracies and non-Gaussian structure in the posterior. The neural network itself does not prevent MCMC sampling of stellar parameters, since the sampling is performed over $\theta$, not over the network weights. However, poorly behaved emulator surfaces (e.g., sharp interpolation artifacts or regions of sparse training coverage) can induce artificial structure in the posterior and must therefore be diagnosed carefully if utilized. 

Interpolation error in observable space does not necessarily translate into degraded precision in inferred stellar parameters (e.g., mass, age, etc.), as the mapping from observables to model parameters is non-linear. Errors in observables that do not strongly influence the stellar parameters will have little effect on the inferred mass or age. In contrast, errors that lie along degeneracy directions, where different parameter combinations produce similar observables, can significantly increase the uncertainty in the inferred parameters. It is therefore insufficient to assess emulator performance solely through residual statistics in observable space.

By projecting output errors onto the principal directions of the Jacobian, we differentiate between benign emulator noise, which occurs in directions the observables are insensitive to, and critical noise that inflates the final uncertainties in $\theta$. If emulator uncertainty remains subdominant to observational uncertainty after propagation, then the emulator fidelity is adequate for the intended application. If not, we quantify the degree to which parameter uncertainties are degraded. This procedure provides a direct, application-specific validation of emulator performance in terms of its impact on stellar characterization.

In practice, mass inference proceeds as follows. As defined in Section~\ref{sec:intro}, the emulator $\tilde{f}$ approximates the true simulator $f$ with a position-dependent interpolation error $\epsilon_{\rm em}(\theta)$. Given a set of seismic observables (here, $\Delta\nu$ and $\nu_{\rm max}$) for a target star, we invert $\tilde{f}$ to find the input mass, $M$, that satisfies
\begin{equation}
    \tilde{f}_{\Delta\nu}(M, \theta_{-M}) = \Delta\nu_{\rm obs}
\end{equation}
where $\theta_{-M}$ denotes the remaining input parameters held fixed at their inferred or assumed values, and $\tilde{f}_{\Delta\nu}$ denotes the $\Delta\nu$ component of $\tilde{f}(\theta)$. The sensitivity of the inferred mass to perturbations in $\Delta\nu$ is governed by the local Jacobian $\partial \tilde{f}_{\Delta\nu}/\partial M$, where by first-order error propagation,
\begin{equation}
    \sigma_M \approx \left| \frac{\partial \tilde{f}_{\Delta\nu}}{\partial M} \right|^{-1} \sigma_{\Delta\nu}
\end{equation}
where $\sigma_{\Delta\nu}^2 = \sigma_{\Delta\nu, {\rm obs}}^2 + \sigma_{\Delta\nu,{\rm em}}^2$ combines observational noise and the standard deviation of $\epsilon_{\rm em}(\theta)$ projected onto the $\Delta\nu$ dimension, the latter estimated from the spread across an ensemble of independently trained FLAX networks. Numerical instabilities in $\tilde{f}$ manifest as local fluctuations in $\partial \tilde{f}_{\Delta\nu}/\partial M$. Where this gradient is anomalously small, even modest $\Delta\nu$ uncertainties are amplified into large mass errors, and spurious local extrema can produce false solutions.

\bibliographystyle{aasjournal}
\bibliography{references}

%%%%%%%%%%%%%%%%%%%%%%%%%%%%%%%%%%%%%%%%%%%%%%%%%%

% Don't change these lines
%\bsp	% typesetting comment
\label{lastpage}
\end{document}